\documentstyle[preprint,aps,version2,eqsecnum]{revtex}
\begin{document}

\begin{title}
Critical Behavior of Charge Density Waves\\
Below Threshold: Numerical and Scaling Analysis
\end{title}

\author{Daniel S. Fisher}
\begin{instit}
Lyman Laboratories, Harvard University, Cambridge MA 02138
\end{instit}
\author{A. Alan Middleton}
\begin{instit}
Physics Department, Syracuse University, Syracuse, NY 13244
\end{instit}
\receipt{}

\begin{abstract}
The critical behavior of pinned charge density waves (CDW's) is
studied as the threshold for sliding is approached.
Using the Fukuyama-Lee-Rice Hamiltonian with relaxational dynamics,
the polarization and linear response are calculated numerically.
Analytic bounds on the subthreshold motion are used to develop
fast numerical algorithms for evolving the CDW configuration.
Two approaches to threshold, ``reversible'' and ``irreversible'' are
studied, which differ in the details of the critical behavior.
On the irreversible approach to threshold, the response due
to avalanches triggered by local instabilities dominates the
polarizability, which diverges in one and two dimensions.
Such ``jumps'' are absent on the reversible approach.
On both the reversible and irreversible approach in
two dimensions, the linear response,
which does not include the jumps, is singular, but does not diverge.
Characteristic diverging length scales are studied using
finite-size scaling of the sample-to-sample
variations of the threshold field in finite systems and
finite-size effects in the linear polarizability and
the irreversible polarization.
A dominant diverging correlation length is found which controls
the threshold field distribution, finite-size effects in the
irreversible polarization, and a cutoff size for the avalanche size
distribution.
This length diverges with an exponent $\nu \approx 2.0, 1.0$ in
dimensions $d=1,2$, respectively.
A distinct exponent describes the finite-size effects for the linear
polarizability in single samples.
Our results are compared with those for related models and
questions are raised concerning
the relationship of the static critical behavior below threshold
to the dynamic critical behavior in the sliding state above
threshold.

\end{abstract}


\section{Introduction}
\label{Introduction}
Despite many years of experimental and theoretical work
\cite{reviews}, much of the behavior
of systems which exhibit sliding charge density waves (CDW's) is still
puzzling.
For most collective properties of these materials, both thermal
fluctuations and defects in the charge density waves (e.g.,
dislocations) appear to play minor roles.
The main problem is thus that of an elastic medium moving through a
random potential caused by impurities.
Such systems are quite ubiquitous \cite{DSFfriction},
arising for weakly pinned vortex
lattices in superconductors and various kinds of driven interfaces
in inhomogeneous media.

The phenomenology of the CDW is based on the existence of two ``phases''.
Above a sharp threshold driving
force, $ F_{T}$, proportional to the electric field, the CDW moves with a
non-zero mean velocity $v$, and the behavior appears to be history
independent.
One of us has proved elsewhere that in this regime there is a unique
periodic steady state \cite{uniqueaam} for CDW models without
dislocations.
Below the threshold force, on the other hand, the CDW relaxes towards
one of many metastable minima and is then stationary at long times
(neglecting slow, thermally activated creep processes
\cite{uniqueaam,dsf:mft,PBLRammal,aam:thermal}).
In this regime, the behavior is strongly hysteretic due to the many
minima.
As threshold is approached from below, sections of the CDW become
unstable and start to slide locally, only to
be stopped by neighboring regions which are more strongly pinned.
These give rise to a non-linear response to changes in $F$.
The cascade of avalanches which occurs bears considerable resemblance
to that found in other systems with collective transport, for example
models of ``sand piles'' and motion of geologic faults
\cite{marg-stable,Carlson-Langer}.
The maximum size of these ``avalanches'' diverges as threshold is approached,
leading eventually to the sliding of the whole system.
It is the properties of CDW's as threshold is approached from below
that will be the main subject of this paper.

\subsection{Model}
\label{model}
The model of CDW's that we study is a simplified version of the
Fukuyama-Lee-Rice Hamiltonian \cite{LeeRice}.
This model focuses on the phases $ \varphi_{ i}$ of
the CDW at impurity sites $i=1,\ldots,N$,
which, for simplicity, are chosen to lie on
a regular linear, square, or cubic lattice of dimension $d$.
Each impurity favors a fixed random phase of the CDW, $ \beta_{ i}$, modulo
$2\pi$, and couples to the phases
with strength $h$, which we take to be uniform.
The effective Hamiltonian is then
\cite{dsf:mft,LeeRice,SCF,periodicobs,miscnumerics}
 \begin{equation}
\label{eq:energy}
{\cal H} = \frac{1}{2} \sum_{ \left( i,j \right)} ( \varphi_{ j}- \varphi_{
i})^2
-h\sum_{i=1}^{N} \cos( \varphi_{ i}- \beta_{ i}) -F(t) \sum_{i=1}^{N}
\varphi_{ i},
 \end{equation}
where the first term represents the elastic interactions between the
CDW at nearest neighbor impurity sites $ \left( i,j \right)$
and the last term represents
the effect of a spatially uniform driving force $F$, which may be a function
of the time $t$.
The equations of motion are purely relaxational
\cite{reviews,LeeRice,SCF,miscnumerics,PBLoned}
 \begin{equation}
\label{eq:motion}
\frac{d \varphi_{ i}}{dt}=-\frac{\partial {\cal H}}{\partial  \varphi_{ i}}=
 \Delta  \varphi_{ i} +h \sin( \varphi_{ i}- \beta_{ i}) +F
 \end{equation}
so that the system just slides down the many dimensional potential
given by  Eq.~(\ref{eq:energy}).
Here $\Delta  \varphi_{ i} \equiv \sum_{\delta} ( \varphi_{ i+ \delta}-
\varphi_{ i})$, with $\delta$ the
nearest neighbor vectors, is the lattice Laplacian.
The preferred phases $\{ \beta_{ i}\}$ are independent, uniformly
distributed, random variables, chosen in the interval $[0,2\pi)$.

Because of the non-linear nature of the equations of motion, very few
analytic results are available: perturbation theory
\cite{SCF,MatsukawaPert} is possible only
for large fields $F \gg  F_{T}$ and some general bounds on the behavior
can be derived (see Sec.~\ref{bounds} and Ref.\ \cite{uniqueaam}).
Mean field theory, valid in the limit of long range elastic
interactions, has been investigated by one of us in some
detail \cite{dsf:mft}, but expansions about mean field theory are
difficult, although very recently progress has been made
above threshold \cite{narayanfisher}.
Since we are interested in the critical behavior near threshold in
finite-dimensional systems, we have thus resorted to extensive
numerical simulations, making use of analytic bounds.
The numerical methods are discussed in  Sec.~\ref{numerics}.

\subsection{Results for polarization for two approaches to threshold}

For the simulations, we use periodic boundary conditions on one, two,
and three dimensional ``cubes'' of volume $L^{d}$.
Starting from an initial configuration $  \varphi_{ i}^{\rm init}$, which is
obtained
after relaxing to a local minimum of ${\cal H}$, we adiabatically
increase (or decrease) the
uniform force $F$, letting the $\{ \varphi_{ i}\}$ relax to a
local minimum for each value of $F$.
One of the primary quantities that we study in
 Sec.~\ref{numericrit} is the polarization (density)
 \begin{equation}
P \equiv L^{-d} \sum_{i} ( \varphi_{ i}-  \varphi_{ i}^{\rm init}).
 \end{equation}

As $F$ is increased, the polarization increases both by continuous motion
--- due to the smooth evolution of the local minimum --- and by discontinuous
jumps which occur when the local minimum of ${\cal H}$, in which the
CDW configuration is, disappears, i.e., a {\em local} saddle
node bifurcation takes place.
Except on the set of measure zero in $F$ for which the jumps occur, we
can define a linear a.c.\ polarizability density as the response to an
infinitesimal additional a.c.\ force $\delta F(\omega)$:
 \begin{equation}
\chi(\omega)\equiv\frac{\delta \left< \varphi(\omega) \right>}{\delta
F(\omega)}
 \end{equation}
where $ \left< \varphi \right> \equiv L^{-d}\sum \varphi_{ i}$.
In the limit of zero frequency, the linear d.c.\ polarizability
$ \chi_{0} \equiv \chi(\omega \rightarrow 0)$ will {\em not}
in general be equal to the derivative of the polarization density,
because of the discontinuous jumps.
Although precursors to the jumps contribute to $ \chi_{0}$, the discontinuous
changes in polarization resulting from the jumps themselves are not
included in $ \chi_{0}$.
As we will see, however, we can define a polarizability for
increasing $F$ by
 \begin{equation}
\label{chidndef}
 \chi^{\uparrow}(F)=\lim_{\delta F  \rightarrow 0^{+}} \lim_{L  \rightarrow
\infty}
\frac{P(F+\delta F)-P(F)}{\delta F},
 \end{equation}
which {\em does} include the jumps.
If $F$ is subsequently decreased, the regions that jumped forward
on increasing $F$ will, because of hysteresis \cite{dsf:mft,miscnumerics},
not jump back in the same fashion, so that
{\em in general},
 \begin{equation}
 \chi^{\downarrow}(F) \neq  \chi^{\uparrow}(F) \neq  \chi_{0}(F) \neq
\chi^{\downarrow}(F),
 \end{equation}
where $ \chi^{\downarrow}(F)$ is defined as is in  Eq.~(\ref{chidndef}), but
with the
limit $\delta F  \rightarrow 0^{-}$.
Nevertheless, there are special system histories in which no jumps
occur and for which all the polarizabilities are equal.

In order to get reproducible results which do not depend on the
initial conditions, we study a particular history: increasing $F$
initially to $ \mbox{$F_{T}^{+}$}$, the threshold force for positive $F$,
then
decreasing the force to the opposite threshold $ \mbox{$F_{T}^{-}$} < 0$, and
then back up to $ \mbox{$F_{T}^{+}$}$.

Since the last local minimum of ${\cal H}$ to disappear as $F$ is
increased is {\em unique} \cite{uniqueaam}
(up to uniform shifts of all phases by a
multiple of $2\pi$), the configuration at $ \mbox{$F_{T}^{+}$}$ (and likewise
at
$ \mbox{$F_{T}^{-}$}$) is unique. After the first increase to $
\mbox{$F_{T}^{+}$}$, the
system can be cycled back and forth to $ \mbox{$F_{T}^{-}$}$.
On the now-uniquely-defined
subsequent increases to $ \mbox{$F_{T}^{+}$}$ (and generically on the initial
increase), the jumps in $P$ become larger as $ \mbox{$F_{T}^{+}$}$ is
approached
as larger regions of the
system reach local thresholds, go unstable, and increase
the elastic forces on their neighboring
regions.
Right at threshold, a local instability leads to motion of the whole
system by $2\pi$.
Thus we anticipate that there should be a correlation length
characterizing the size of the ``avalanches''
which diverges as $F \nearrow  \mbox{$F_{T}^{+}$}$.

Concomitantly, the polarizability $ \chi^{\uparrow}$ diverges, on the
{\em irreversible approach} to threshold, with a divergence of the
form
 \begin{equation} \label{gammadef}
 \chi^{\uparrow}(F) \sim (F_{T}-F)^{-\gamma}.
 \end{equation}
The polarization itself will also diverge if $\gamma > 1$.
The d.c.\ linear polarizability, $ \chi_{0}$, will be strongly dependent on
$F$ with this approach to threshold, but as shown below in
 Sec.~\ref{numericrit}, it will
follow a curve which is {\em smooth} almost everywhere
in the limit $L  \rightarrow \infty$ and which is distinct from $
\chi^{\uparrow}$.
Equivalently, one can calculate the configuration averaged
$\overline{ \chi_{0}(F)}$, which does not include the effects of the jumps.
The precursors to jumps will not contribute significantly to this mean.
As $F \nearrow  \mbox{$F_{T}^{+}$}$, we find in the limit of an infinite
system
 \begin{equation}\label{gammalidef}
 \chi_{0}(F) \approx  \chi_{T} -  A^{I} ( \mbox{$F_{T}^{+}$}-F)^{-
\gamma_{\ell}^{I}}
 \end{equation}
with the exponent $ \gamma_{\ell}^{I} <0$, so that $ \chi_{0}(F)$ exhibits
only an
upwards cusp to a constant value $ \chi_{T}$ at threshold.
(Note that in any finite system $ \chi_{0}(F)$ diverges as
$( \mbox{$F_{T}^{+}$}-F)^{-1/2}$; but the amplitude of this divergence is
negligible in a large system. It dominates only very near to threshold.)
In two dimensions, which we have studied most extensively, we find
$\gamma= 1.8\pm 0.15$ and
$ \gamma_{\ell}^{I}=  -0.40\pm 0.12$.

When the force is decreased from $ \mbox{$F_{T}^{+}$}$, initially no jumps
occur
and the minimum of ${\cal H}$ evolves smoothly.
In this regime, which appears to persist for a finite range of $F$ in
large systems, the evolution is reversible, so that $ \chi^{\uparrow}=
\chi^{\downarrow}= \chi_{0}$.
For this history,
the field can be increased back to threshold, yielding a {\em
reversible approach} characterized by a {\em cusp}
 \begin{equation}\label{gammalrdef}
 \chi_{0}(F) \approx  \chi_{T} -  A^{R} ( \mbox{$F_{T}^{+}$}-F)^{-
\gamma_{\ell}^{R}}
 \end{equation}
with $ \gamma_{\ell}^{R}= -0.42\pm 0.05$ in two-dimensional systems.
Within our error bars, $ \gamma_{\ell}^{I}= \gamma_{\ell}^{R}$ (although the
amplitudes
$ A^{I}$ and $ A^{R}$ differ), which is somewhat surprising in light of the
large differences between the two approaches: in the irreversible
approach, the linear polarizability is only a small addition to a much
larger polarizability $ \chi^{\uparrow}$ dominated by the jumps.
We propose in  Sec.~\ref{numericrit}
that the scaling {\em form} for the distribution of
linear eigenmodes is independent of the approach to threshold,
with a common frequency scale $( \mbox{$F_{T}^{+}$}-F)^{\mu}$, where
$\mu \approx 0.50$, but with history-dependent scaling functions.

\subsection{Finite size effects and avalanche size distribution}

In order to investigate the correlation lengths which characterize the
critical behavior near threshold, we study finite size effects in some
detail in  Sec.~\ref{sec:fss} and determine the avalanche size distribution
for the irreversible approach to threshold.
In general, one expects the singular properties of large finite size
systems to exhibit finite-size scaling behavior as functions of
$L/\xi$, with $\xi$ the correlation length.
For example, the polarization density on the irreversible approach to
threshold is expected to behave as
 \begin{equation}
P^{\uparrow}_{\rm sing}(F,L) \approx ( \mbox{$F_{T}^{+}$}-F)^{-\gamma}
\Phi^{\uparrow}(L/\xi).
 \end{equation}
Here, however, because of the randomness and existence of a threshold
in finite systems, care must be taken to use the appropriate sample
specific finite system quantities, as we see in  Sec.~\ref{sec:fss}.

{}From the polarizability $ \chi^{\uparrow}$ on the irreversible approach, as
well as the width of the distribution of threshold fields, we find a
correlation length
 \begin{equation}
\xi \sim ( \mbox{$F_{T}^{+}$}-F)^{-\nu}
 \end{equation}
with $\nu =  2.01 \pm  0.02, 1.01 \pm  0.03$
in $d=1,2$ respectively.
To within our errors, $\nu$ seems to saturate the inequality $\nu
\geq 2/d$ \cite{chayesetal}.

The divergence of $\xi$ is in agreement with the divergence
of the measured characteristic size of avalanches as threshold is
approached.
At threshold, the distribution of avalanche sizes is
scale-invariant, with a distribution quite similar to that seen
for sand-pile models of the same dimensionality ($d=2$).
The CDW model is
clearly not ``self-organized'', as the scale-invariant
behavior is seen only at threshold.
The apparent connection between the two models is quite
interesting, though, and it is briefly developed below.

One can also investigate finite size corrections to the polarizability
in the reversible approach to threshold.
Surprisingly, these are characterized by a distinct characteristic
length
 \begin{equation}\label{nuldef}
 \xi_{l} \sim ( \mbox{$F_{T}^{+}$}-F)^{- \nu_{\ell}}
 \end{equation}
with $ \nu_{\ell} \approx 0.44 \pm 0.05 <
\nu$ in $d=2$.
The origin of this second length, which naively violates the
inequality for $\nu$, is quite subtle.
It is connected to the effects of the smooth potential and the absence
of a natural connection (such as a magnetic field in conventional
equilibrium phase transitions) which links one side of the transition
to the other.
If the polarizability in the reversible approach is scaled with the
dominant correlation length $\xi$, there will be {\em no finite size
corrections to $ \chi_{0}$!}

In the last section of this paper, we discuss the finite size lengths
and related issues, consider some possible scaling laws, and raise
questions for future work. A summary of our numerical results for
critical exponents is presented in Table \ref{table1}. The error
bars for the exponents are subjective, except for $ \nu_{\rm T}$, where the
error bars are statistical, and reflect the range of
values which are consistent with the data in the appropriate
scaling regime (see figures and discussion in each section).

\subsection{Related models}

Before proceeding with a more detailed discussion of the CDW model of
 Eq.~(\ref{eq:motion}), we first define several other models to which we
compare some of our results.

In the limit that the range of interactions in  Eq.~(\ref{eq:motion})
becomes infinite range, mean field theory becomes valid, and one can
replace $z^{-1}\sum_{\delta} \varphi_{ i+ \delta}$ (with $z=2d$ the
coordination
number),
by a self-consistently calculated mean field $\overline{\phi}(t)$,
so that the $ \Delta  \varphi_{ i}$ term of  Eq.~(\ref{eq:motion}) is
replaced by
$\overline{\phi}(t)- \varphi_{ i}$.
This model has been studied in some detail in Ref.\ \cite{dsf:mft}
and shows some
features qualitatively similar to the present finite dimensional
results, particularly the presence of reversible and irreversible
approaches to threshold and the spectrum of local soft modes.

A one dimensional incommensurate version of  Eq.~(\ref{eq:motion}) has also
been studied \cite{SNCandDSF}.
Here the interactions are nearest neighbor, but the pinning phases
$ \{ \beta_{ i}\}$ are chosen to be quasiperiodic (rather than random), i.e.,
$ \beta_{ i}=2\pi\alpha i$, with $\alpha$ an irrational, usually chosen to
be the golden mean.
For $h$ greater than a critical value $h_{c}$, this system exhibits a
non-zero threshold somewhat similar to the random system.

Finally, several authors have studied \cite{Mihalyetal,Webman87}
a simple ``random friction''
model, in which the cosine pinning potential is
(essentially) replaced by a periodic sawtooth with
 \begin{equation}
 V_{ i}( \varphi_{ i})= h_{ i} \varphi_{ i}\ \ \  {\rm for}\ \ \   \beta_{
i}> \varphi_{ i}>2\pi- \beta_{ i},
 \end{equation}
and $ h_{ i}$ random with some distribution; $ V_{ i}$ is periodic with
period
$2\pi$.
The discontinuity in $ V_{ i}$ at $ \beta_{ i}$ does not affect the steady
state dynamics of the moving phase, but only stops the phases from
``backsliding'' below threshold.
At threshold, the distortion of the phases in this model can be
calculated directly, as can the distribution of threshold fields.
The threshold field $F_{T}(\{ h_{ i}, \beta_{ i}\})$ is simply the average of
$ h_{ i}$ over the system, so that in a system of size $L^{d}$, the width
of the distribution of threshold fields is $ \mbox{$\Delta F_{T}(L)$} \sim
L^{-d/2}$.
The mean distortions at threshold (and for all fields above threshold)
behave as
 \begin{equation}
\overline{( \varphi_{ i}- \varphi_{ j})^{2}} \sim  \left| i-j
\right|^{(4-d)/2}
 \end{equation}
for $d<4$, since the discontinuities in the potential do not play a role
(we note that a ``ratcheted kick'' model \cite{uniqueaam}, where the
phase is advanced by a finite amount when it reaches a discontinuity, has
much more complicated behavior above threshold; see
 Sec.~\ref{sec:discussion}).

\section{Analytic bounds}
\label{bounds}

The behavior of CDW's below threshold is characterized, as discussed
above, by many locally stable configurations and concomitant
hysteresis.
Above threshold, the many non-linearly interacting degrees of freedom
might be expected to lead to strong sensitivity of the motion to
initial conditions, aperiodicity, or non-uniqueness.
As one of us has shown elsewhere \cite{uniqueaam},
however, the convexity of the
interactions between the phases assures that at long times, CDW's
above threshold approach a {\em unique periodic} steady state (as seen
in numerical simulations \cite{periodicobs,miscnumerics,PBLoned}).
In this section, we show that similar convexity arguments lead to a
partial ordering of the configurations below threshold, bounds on the
motion, and the uniqueness of the configuration {\em at} threshold.

\subsection{The no-passing rule}
\label{sec:nopassing}
The central result upon which these conclusions are based is what we
call the ``no-passing'' rule.
This rule severely restricts the
behavior of solutions to the equation of motion.
Suppose one has two solutions to the equation of
motion  Eq.~(\ref{eq:motion}), for the same realization of the pinning and
same drive field.
If the initial conditions are such that the values of all the phases for
one solution (the ``greater'' solution)
exceeds the phases in the other solution
(the ``lesser'' solution) at each point in space,
the greater solution can never
be ``passed'' by the lesser one (see  Fig.~\ref{nopassing}).
Physically, if the phases of the two solutions approach each other at some
site,
the pinning and drive forces cancel, but the elastic forces due
to neighboring sites keep the
phases from passing through each other, since
the elastic forces tend to flatten out the configuration.
This rule, though quite simple and easily derived, provides a partial
ordering
for the stationary
solutions to the equations of motion and directly implies that
the velocity is a unique function of the applied field.

We now justify this rule in more detail.
Consider two solutions to the equations of motion
Eq.~\ref{eq:motion}, $\{  \varphi_{ i}^{1}(t)\}$ and $\{  \varphi_{
i}^{2}(t)\}$,
for the same realization of the disorder $\{ \beta_{ i}\}$,
with initial conditions
chosen so that $  \varphi_{ i}^{1}(0) <   \varphi_{ i}^{2}(0)$, for all $i$.
We will say that such a configuration $\{  \varphi_{ i}^{1}(t)\}$ is {\em
less than}
$\{  \varphi_{ i}^{2}(t)\}$ at $t=0$.
Each set of phases is driven by the same, possibly time dependent,
external field, $F(t)$.
Define the differences
 \begin{equation}
 \epsilon_{ i}(t)=  \varphi_{ i}^{2}(t)-  \varphi_{ i}^{1}(t).
 \end{equation}
Subtracting the equations of motion  Eq.~(\ref{eq:motion}) for the
two solutions gives
 \begin{equation}\label{eq:epsi}
\frac{d \epsilon_{ i}(t)}{dt} =  \Delta \epsilon_{ i}(t) + h \left[ \sin(
\varphi_{ i}^{1}+ \epsilon_{ i}- \beta_{ i})
-\sin(  \varphi_{ i}^{1}- \beta_{ i}) \right].
 \end{equation}
Let $ j(t)$ be a site where $\epsilon$ is equal to its minimum
value.
Since $ \Delta\epsilon_{ j(t)}$ is non-negative
and the second term of  Eq.~(\ref{eq:epsi}) is
bounded in magnitude by $h \left|  \epsilon_{ i} \right|$,
 \begin{equation}
\frac{d}{dt}\epsilon_{ j(t)}>-h\epsilon_{ j(t)}(t).
 \end{equation}
It follows that $\epsilon_{ j(t)}(t)$ decays no faster than
exponentially to zero (indeed it will often increase)
and therefore, for all $ i$, $  \varphi_{ i}^{1}$ cannot coincide
with or cross $  \varphi_{ i}^{2}$ at any time
(in the sliding state, $\min_{ i}( \epsilon_{ i})$ is bounded below by a
constant
which depends on the initial configurations $  \varphi_{ i}^{1},  \varphi_{
i}^{2}$ \cite{uniqueaam}).
Thus we conclude that if $\{  \varphi_{ i}^{1}(0)\}$ is less than $\{
\varphi_{ i}^{2}(0)\}$,
$\{  \varphi_{ i}^{1}(t)\}$ will be less than $\{  \varphi_{ i}^{2}(t)\}$,
for all $t>0$.

It is clear that this no-passing rule relies crucially on
the elastic potential between sites being convex.
Note that for models
where the elastic potential is not convex, this result
does not hold, and, in fact, many of the conclusions that we
will derive here do {\em not} hold for models with phase
slip \cite{phaseslip}.

\subsection{Consequences of no-passing}

The no-passing rule has several immediate, useful consequences.
The first is the uniqueness of the velocity.
Suppose there are two solutions to the equations of motion in a
finite system, for the
same field $F(t)$ and pinning realization $ \{ \beta_{ i}\}$.
By discrete translation
invariance of the equations of motion by multiples of $2\pi$, either
of the two solutions initially can be translated to become a lesser solution.
By the no-passing rule, the average velocity of the lesser solution is
bounded above by that of the greater solution.
Since the choice of initially lesser configuration is arbitrary,
all solutions to the equations of motion in finite systems
must have the same velocity, in the limit of long times.

This immediately implies that the threshold field is unique, since
moving solutions cannot coexist with stationary solutions.
The threshold configuration itself is almost always unique.
Since the pinning potential is random, generically no
more than one minimum will disappear simultaneously at $ F_{T}$.
The last stationary solution to disappear as $F$ increases is thus,
with probability one, the unique threshold configuration, modulo
uniform $2\pi$ shifts of all the phases.

Another important consequence of the no-passing rule
is the bounding of the motion for fields $F$ in
the static range, $ \mbox{$F_{T}^{-}$} \leq F \leq  \mbox{$F_{T}^{+}$}$.
This bound shows that, given an initial configuration and for monotonic
changes in the field, the final
configuration approached depends only on the final
field.
In particular, the final configuration is
independent of the rate of change of the applied field.
This allows for a natural ordering of the static states which is useful for
understanding hysteresis and for the simulation of the model (see
 Sec.~\ref{numerics}).

Consider any particular realization of the pinning $\{ \beta_{ i}\}$.
Let $\{ \varphi_{ i}(t)\}$ be a solution to the equations of motion
Eq.~(\ref{eq:motion})
for a (possibly) time-dependent field $F(t)$ which is bounded above by a
constant $F^{*}$, with
 \begin{equation}
F(t) \leq F^{*} \leq  \mbox{$F_{T}^{+}$},
 \end{equation}
for all $t$.
Define ${\cal A}^{*}(\{ \varphi_{ i}(0)\})$ as the set of all
configurations $\{ \varphi_{ i}^{*}\}$ that are stationary
at the field $F^{*}$ and that are greater
than the initial configuration, i.e.,
$ \varphi_{ i}(0) \leq  \varphi_{ i}^{*}$, for all $i$.
Suppose that the configuration $\{ \varphi_{ i}^{*}\}$ is a member of ${\cal
A}^{*}$.
The ``no passing'' rule implies that $ \varphi_{ i}(t) \leq  \varphi_{
i}^{*}$,
for all $t>0$ and all $i$;
as the initial configuration evolves, it cannot pass
any configuration $\{ \varphi_{ i}^{*}\}$
that is both stationary at the bounding field $F^{*}$ and greater than
the initial configuration.
 Fig.~\ref{leastconfig} schematically
shows some of the configurations $\{ \varphi_{ i}^{*}\} \in {\cal A}^{*}$.
If the field $F(t)$ is non-decreasing in time,
with $F^{*}=\lim_{t  \rightarrow \infty} F(t)$, then
$\{ \varphi_{ i}(t)\}$ approaches a stationary configuration defined by
$ \varphi_{ i}^{\infty} = \lim_{t  \rightarrow \infty}  \varphi_{ i}(t)$.
This limit
configuration is stationary at field $F^{*}$ and is greater than the initial
configuration $\{ \varphi_{ i}(0)\}$, therefore it belongs in
the set ${\cal A}^{*}$.
The configuration $ \varphi_{ i}^{\infty}$ is the
{\em least} configuration that is greater than $\{ \varphi_{ i}(0)\}$
and stationary at $F^{*}$: it has the
property that $ \varphi_{ i}^{\infty} \leq  \varphi_{ i}^{*}$ for all $\{
\varphi_{ i}^{*}\} \in {\cal
A}^{*}$, since the configuration $\{ \varphi_{ i}(t)\}$ cannot pass
{\em any} of the configurations in ${\cal A}^{*}$.
For non-decreasing $F(t)$, the final configuration approached is thus
{\em unique} and {\em independent} of the rate at which $F(t)$ approaches
$F^{*}$.
The final configuration
depends only on the initial configuration and the value of $F^{*}$.
As we see below,
this result enables the stationary configurations which occur for
adiabatically changing $F$ to be computed more efficiently.

\subsection{Other physical systems}

We note that the no-passing rule and its consequences are applicable
to models of other physical systems, including the motion of
interfaces in random systems, e.g., fluid interfaces in porous media
\cite{fluids}.
The two most important requirements for the no-passing rule are
convex elastic interactions and the parameterization of distortions by
a {\em single} field defined on the elastic medium.
Flux flow in superconducting films, for example, fails both of these tests:
the flux lattice can rearrange itself plastically under large strains
(non-convex elastic forces) and there are two internal coordinates
(the position vector of individual fluxoids).
It has been shown \cite{phaseslip}
that models with non-convex elastic interactions can have hysteresis in the
velocity vs.~field relation.
Note that is even possible to define {\em zero-dimensional}
models that have two (internal) coordinates where the velocity is
hysteretic  \cite{twocoords}.

\section{Numerical algorithms}
\label{numerics}

Previous simulations \cite{periodicobs,miscnumerics,Mihalyetal,SibaniPBL}
of the lattice CDW model have used direct
numerical integration of the equations of motion.
Some of our results were also obtained with this method.
When we used a second-order predictor-corrector
method to integrate the equations of motion,
we found that time steps in the range 0.05-0.1 were sufficient for
finding stationary configurations (and also for simulating the sliding
state).
For the parameters we used, decreasing the time step did not change the
results.  A great disadvantage of this method is that the relaxation time to
a
static configuration can be quite
long.  This occurs when the simulation is converging towards a configuration
that has soft modes (i.e., linear relaxational modes with relaxation
rates $\ll h$).
As we shall see, such soft modes are often important.
This problem, not surprisingly, is especially bothersome near threshold.

For fields in the static range, of primary interest here,
we used an alternative method to
find static configurations and to calculate the changes in configuration that
occur when the drive field is changed.
It is similar to a method previously used in the incommensurate
case  \cite{SNCandDSF}.
This method, which relies on the
existence of the no-passing rule for the CDW model, is often over
two orders of magnitude faster than direct numerical integration.
In fact, we found this algorithm essential to achieve well-converged
configurations in large systems for fields near threshold.

For an example of the method used, suppose that, using numerical
integration, one has obtained an initial configuration stationary at a
field $F^{0}$, and one wishes to find the configuration that is static
at some greater field $F^{*}$, $F^{0} < F^{*} <  \mbox{$F_{T}^{+}$}$,
that would result from integrating
the equations of motion, with field $F^{*}$ for times $t>0$.
This configuration is the unique {\em lowest} configuration,
above the initial configuration, that is static at field
$F^{*}$, by the above discussion of the ordering of static states.
Any algorithm that determines this lowest configuration will be an acceptable
method.
The method that we use is to advance each degree of freedom towards,
{\em but not beyond}, the nearest local minimum of the local energy
defined at each site $ i$,
${\cal H}_ i( \varphi_{ i})$, which is determined by calculating
the total energy ${\cal H}$ with all neighboring phases {\em fixed}.

To advance individual phases toward the minima of ${\cal H}_ i$, we
use a variant
of the Newton-Raphson algorithm, which we modify to avoid overshooting
the minima, while retaining rapid convergence.  We need to find
zeroes of the velocity
 \begin{equation}
 \dot{\varphi}_{ i}=-\partial{\cal H}/\partial{ \varphi_{ i}}\equiv -{\cal
H}'_{ i}
\equiv F_{ i} + V'_{ i}-2d \varphi_{ i},
 \end{equation}
where the local field $F_{ i}\equiv \sum_{\delta} \varphi_{ i+ \delta}+F$
and the pinning force $V'_{ i}\equiv h\sin( \varphi_{ i}- \beta_{ i})$.
Instead of the usual iterative map
$ \varphi_{ i}  \rightarrow  \varphi_{ i} - {\cal H}'_{ i}/{\cal H}''_{ i}$,
we solve for zeroes of $ \dot{\varphi}_{ i}$ by the map
 \begin{equation}
\label{modNR}
 \varphi_{ i}  \rightarrow  \varphi_{ i} -
{\cal H}'_{ i}/({\cal H}''_{ i}-c{\cal H}'_{ i}),
 \end{equation}
where $c$ is a constant that depends on the pinning strength \cite{footc}.
After a single iteration of this map for each site,
the local fields are updated.
At each step, the phases only increase, and the local fields
$F_{ i}$ increase.
This map is iterated until a fixed configuration
is reached, which corresponds to a stationary configuration.
Since the phases never ``pass'' a minimum of the energy, the configuration
reached is the lowest configuration, greater than the initial configuration,
that is static at field $F^{*}$.
The fixed configuration obtained by this method is therefore the same as
would be obtained by numerical integration of the equations of motion.

The threshold fields for each particular realization of the pinning was
found by bisection.
Upper and lower bounds were found for the threshold field (estimated from
previous runs or taken to be given by
$0 <  \left|  F_{T} \right| < h$), and a
configuration stationary at the lower bound was found.
These bounds were then
improved by determining whether the configuration stationary at the
lower field could be evolved, by the methods just described, into a
configuration that was stationary at a trial field equal to the mean of the
two
bounds.  If so, then the lower bound and static configuration were updated.
If not, the upper bound was lowered to the trial field.
Note that the no-passing rule implies that if the movement of $all$
of the phases from their initial positions becomes greater than
$2\pi$ at any iteration, the applied field must be greater than the
threshold field.
This provides an unambiguous
criterion to determine when the threshold field has been exceeded.

The computations were carried out on the 16K CM-2 Connection
Machine at
Argonne National Laboratories and the CM-2 at Syracuse University.
For small systems,
many realizations of the pinning were studied simultaneously.
For all simulations, we used periodic boundary conditions.
The pinning strengths $h_{ i}$ were taken to be uniform, as in
 Eq.~(\ref{eq:energy}), with
values of 2.5, 5.0, and 7.5 in one, two and three dimensions,
respectively. These values of the pinning strength, $h=(2.5)d$,
were chosen to yield a
Lee-Rice length $\xi_{LR}$, the scale
at which the elastic and pinning energies are comparable \cite{LeeRice},
of approximately one lattice unit.
Since regions of volume $\xi_{LR}^{d}$ act effectively as single degrees of
freedom,
this choice allows for the most efficient
simulation of many effective degrees of freedom.
With these values for the pinning strength, we find the threshold fields for
large systems to be $1.338 \pm 0.004$, $1.490 \pm 0.005$, and $1.282 \pm
0.002$ in one,  two and three dimensions, respectively.
That these are of order unity is consistent
with the Lee-Rice length being approximately one lattice unit.
(These values of the pinning
strengths are also very similar to those used in Ref.\ \cite{SibaniPBL}.)

\section{Numerical results for critical behavior}
\label{numericrit}

In this section, we present our numerical results for the critical
behavior as threshold is approached from below.
As discussed in the Introduction, the behavior is strongly history
dependent below threshold (this is already seen in the mean field
limit \cite{dsf:mft}).
We thus choose two distinct, well-defined approaches to threshold on
which much of the behavior differs {\em qualitatively}.
Nevertheless, some of the properties are quantitatively similar for these two
histories, suggesting some underlying genericity.
The primary quantities we study are the polarization $P$ and various
polarizabilities as defined in the Introduction.

As discussed above, the configurations at the positive and negative
thresholds, $ \mbox{$F_{T}^{+}$}$ and $ \mbox{$F_{T}^{-}$}$, are unique.
Thus natural reproducible histories can be analyzed in which the field
is slowly swept back and forth from $ \mbox{$F_{T}^{-}$}$ to $
\mbox{$F_{T}^{+}$}$.
The paths on approaching $ \mbox{$F_{T}^{+}$}$ and going away from $
\mbox{$F_{T}^{+}$}$ are,
as we shall see, quite distinct.
If $F$ is increased to above threshold, and then decreased again
slowly, the same configurations will be passed through below $
\mbox{$F_{T}^{+}$}$,
up to uniform $2\pi$ translations.

\subsection{Hysteresis and qualitative behavior}

We now examine the results of a simulation of the histories, for which
the initial configuration is the configuration static at $
\mbox{$F_{T}^{-}$}$.
The applied field is then varied adiabatically;
first increasing to $ \mbox{$F_{T}^{+}$}$, then decreasing back to $
\mbox{$F_{T}^{-}$}$.
 Fig.~\ref{hystloop} shows the polarization as a function of applied field
for a
single two-dimensional system of linear size $L=64$, for this history.
A similar hysteresis loop has been seen previously in simulations of the
random-friction model in one dimension \cite{Mihalyetal}.
We use this loop to uniquely (up to translations by multiples of $2\pi$)
define two paths in configuration space, which are parameterized by the
applied field:
$\{  \varphi_{ i}^{\uparrow}(F)\}$, for field increasing from $
\mbox{$F_{T}^{-}$}$, and
$\{  \varphi_{ i}^{\downarrow}(F)\}$, for field decreasing from $
\mbox{$F_{T}^{+}$}$.

We refer to these histories as the two ``extremal'' histories.
For any system subject to a time dependent field $ \mbox{$F_{T}^{-}$} \leq
F(t) \leq
 \mbox{$F_{T}^{+}$}$, with an initial configuration belonging to an extremal
history, the
evolution of the configuration is bounded by this hysteresis loop:
 \begin{equation}
\label{loopbound}
  \varphi_{ i}^{\uparrow}(F(t)) \leq  \varphi_{ i}(t) \leq   \varphi_{
i}^{\downarrow}(F(t)),
 \end{equation}
for all $ i$ and $t>0$.
This result follows directly from the no-passing
rule: it bounds the changes in polarization for such an initial
configuration.
General time-dependent configurations need not be bounded
by this simple loop.
However, if the applied field changes adiabatically and equals the
threshold field at any time,
the evolution at later times will be bounded by this loop (or uniform
$2\pi$ translations of it.)

For our choice of pinning strength,
we find that the initial section of each of
the extremal paths, where the field is {\em reduced in magnitude} from its
threshold value, $ \mbox{$F_{T}^{\pm}$}$, is {\em reversible}.
For all quasistatic field histories that start with a
configuration that is static at field $ \mbox{$F_{T}^{+}$}$ or $
\mbox{$F_{T}^{-}$}$,
the configuration is a unique function of the field for a range of fields
near
threshold, even if the direction of the field change is reversed.
This is consistent with our observation that, over this field range, {\em no
jumps in the phase occur}, since no local minima vanish.
The polarization is
a smooth function of field over this range,
even for finite systems.
In two dimensions, we find the
range of fields $ \mbox{$F_{T}^{+}$} \geq F \geq  F_{R}$ over which the
system
is reversible is given by $ \mbox{$F_{T}^{+}$}- F_{R}=0.80 \pm 0.03$, while
$ \left|  F_{T} \right|=1.490\pm0.005$.

This reversibility over some range is consistent with that
seen in experiments on CDW's \cite{Ong-hyst}, where the resistance
is measured as a function of the history of the applied field.
The CDW configuration affects the electrical transport properties
of the normal carriers, even when the CDW is pinned.
The resistance in the pinned state is therefore a useful probe
of the history-dependence of the CDW configuration, though the
exact correspondence between the resistance and CDW configuration
is not clear. Duggan, et.\ al., \cite{Ong-hyst} find that when
the field is lowered from above threshold to some distance below
threshold, there is a region
where the resistance is a reversible function of the field. When
the field is lowered further below threshold, hysteresis is evident,
however.
Qualitatively, these results are in agreement with our numerical results
and with mean-field theory \cite{dsf:mft}.

We have examined the critical behavior for the two approaches to the
threshold value of the field, $ \mbox{$F_{T}^{+}$}$, which are defined by the
two
extremal paths, $\{  \varphi_{ i}^{\uparrow}(F)\}$ and $\{  \varphi_{
i}^{\downarrow}(F)\}$.
One history is the {\em irreversible} approach, for which the initial
configuration is the one static at $ \mbox{$F_{T}^{-}$}$.
The critical behavior is given by the behavior of the configurations
$\{  \varphi_{ i}^{\uparrow}(F)\}$ as $F \nearrow  \mbox{$F_{T}^{+}$}$ and is
equivalent to the critical
behavior that would be seen by sweeping the field adiabatically from large
negative field towards $ \mbox{$F_{T}^{+}$}$.
The {\em reversible} history uses the configurations which belong to
the path obtained by lowering the field from $ \mbox{$F_{T}^{+}$}$ (but not
as far
as $ F_{R}$) and then approaching $ \mbox{$F_{T}^{+}$}$ again.
The critical behavior
for this history is given by the properties of the configurations
$\{  \varphi_{ i}^{\downarrow}(F)\}$ near $ \mbox{$F_{T}^{+}$}$ and is
statistically equivalent (by the
{\em statistical}
$\varphi  \rightarrow -\varphi$ symmetry) to increasing the field slowly from
large
negative fields and studying the behavior near $ \mbox{$F_{T}^{-}$}$.

\subsection{Polarization at threshold}

An obvious first question concerning the critical behavior below threshold
is whether the polarization
diverges as the threshold field is approached (for some generic field
history, such as the extremal histories) in an infinite-size system.
In the infinite-range model, if the range of pinning values is bounded, the
polarization cannot diverge \cite{dsf:mft}.
But in an infinite system in finite-dimensions,
there is no constraint that prevents the
polarization from diverging.
However, as shown in \cite{uniqueaam},
there is a strict
bound on the width $W(L)\equiv \max_{i} \varphi_{ i}-\min_{i} \varphi_{ i}$
of a configuration in a system of size $L$, if $h$ is bounded by a
constant $ h_{\rm max}$.

This bound on the width, $W(L) \leq  h_{\rm max} L^{2}/2$, gives a bound for
the
polarization of the threshold configuration:
by the no-passing rule, some
phase must move by less than $2\pi$ for fields below threshold.
The largest amount that a
single phase can increase below threshold is then $2W+2\pi$, since no phase
may differ by more than $W$ from another.
The polarization of a configuration, relative to any static initial
configuration, is therefore bounded by $2\pi+ h_{\rm max} L^{2}$.
As we now discuss, numerical calculations show that the {\em typical}
polarization increases with $L$ much less rapidly than this strict bound.

 Fig.~\ref{threshPvsize} shows our numerical results for $P_{T}$,
the polarization at the threshold field $ \mbox{$F_{T}^{+}$}$, as a function
of $L$,
the linear size of the system, in one and two dimensions.
The initial reference configuration is the one static at $F=
\mbox{$F_{T}^{-}$}$.
Clearly the polarization diverges with system size in both one and two
dimensions, but more slowly than $L^{2}$.
At threshold, the only important length scale is expected to be $L$,
so the simplest behavior that might be expected
for the polarization at threshold, $P_{T}(L)$, is
 \begin{equation}\label{rhodef}
P_{T}(L) \sim L^{\rho}
 \end{equation}
 for some exponent $\rho$.
{}From a fit to the data shown, we estimate $\rho$ to be
$ 1.3\pm 0.3$ in one dimension and
$ 0.8\pm 0.2$ in two dimensions.
We discuss this result in terms of finite size scaling in
 Sec.~\ref{sec:fss}.
The variation in $P_{T}$ from sample to sample appears to be of the
{\em same order} as $P_{T}$, so it would
require many samples at large sizes to obtain a precise estimate for
the exponent $\rho$ (numerically, the r.m.s.\ fluctuations in $P_{T}$
are approximately $0.25P_{T}$ in two dimensions).

\subsection{Subthreshold polarization and linear response}

We now consider the critical behavior of the polarization as threshold is
approached with $F \nearrow  \mbox{$F_{T}^{+}$}$.
On the irreversible approach,
the polarization changes by non-linear jumps, where sections of the
CDW move forward in response to infinitesimal changes in the applied field,
and also a linear response.
Though the non-linear response is very different for
the two approaches to threshold, the linear response appears to have
universal features.

\subsubsection{Definition of reduced field}

In contrast to thermodynamic transitions that occur at a critical
temperature,
for which fluctuations make the definition of the exact location
of the transition temperature ill-defined in a finite size system,
the dynamical system that describes CDW's at zero temperature
has no noise and any finite system has a well-defined threshold,
since the steady-state velocity is either zero or non-zero.
We thus define the reduced field $f$ relative to the threshold
fields for {\em each realization} of the pinning:
 \begin{equation}
\label{def-f}
f \equiv
  2\frac{F -  \mbox{$F_{T}^{+}$}(h,\{ \beta_{ i}\},L)}
        { \mbox{$F_{T}^{+}$} (h,\{ \beta_{ i}\},L)- \mbox{$F_{T}^{-}$}(h,\{
\beta_{ i}\},L)},
 \end{equation}
where the threshold field $ \mbox{$F_{T}^{+}$}$ for a system of size $L$
depends on the
pinning strength $h$ and the realization of the pinning phases $ \{ \beta_{
i}\}$.
For all realizations of the pinning, then, $-2 < f< 0$ in the stationary
phase
and $f>0$ or $f<-2$ in the sliding state.
The reduced field $f=0$ strictly separates the sliding
from the stationary state in each sample.
This definition is necessary when averaging over many samples,
since quantities such as the polarization are not defined in the sliding
state, above threshold.

\subsubsection{Irreversible approach: total response}

Using this definition of the reduced field $f$, we have determined the
polarization $P(f)$ for the irreversible approach to the threshold field
$ \mbox{$F_{T}^{+}$}$ in one and two dimensions and the reversible approach
for
two-dimensional systems.
Figs.~\ref{pol_one} and \ref{pol_two} display the results in one
and two dimensions for various system sizes, for the irreversible
history $\{  \varphi_{ i}^{\uparrow}(F)\}$.
The polarizations are measured relative to that of the configuration
static at the field $ \mbox{$F_{T}^{-}$}$.
In  Fig.~\ref{chi-two}, we plot the numerical derivatives of the polarization
for two-dimensional systems on a log-log scale.
We define $ \chi^{\uparrow}=dP/dF$ as this polarizability,
for the approach to threshold with {\em increasing} field.
{}From this plot of $ \chi^{\uparrow}$ for the largest samples studied, we
deduce the
exponent $\gamma=  1.8 \pm  0.15$
in two dimensions, with $ \chi^{\uparrow} \sim f^{-\gamma}$ for small $f$.
For one-dimensional systems, we calculate $\gamma$ from a fit to $P(f)$;
the polarization diverges as $P \sim f^{-\gamma+1}$. We find $\gamma =
 3.0 \pm  0.5$ in one dimension.

In three dimensions, we have not been able to determine whether the
polarization diverges as $f \nearrow 0$, as the simulation of large
three-dimensional systems requires very large amounts of computer
time.
We display the polarization in  Fig.~\ref{polthreed}
for a $64^{3}$ and a $128^{3}$ system for an approach to threshold
where we have used an initial configuration found by relaxing a configuration
$ \varphi_{ i}\equiv0$ at $F=0$, i.e., slightly different from the $
\varphi_{ i}^{\uparrow}$
approach, giving an initial polarization $P\approx 0$.
The polarization may
be divergent in infinite systems, but only very slowly:
for the $128^{3}$
system, the polarization exceeds $2\pi$ only for fields
within $\sim 0.1\%$ of threshold.
This is in qualitative agreement with
experimental results, where the CDW polarization is less than a wavelength
of the CDW for fields approaching the threshold
value \cite{NMR}.

\subsubsection{Calculation of the linear response}

As the field is increased adiabatically in a finite size system,
the evolution of the CDW configuration is composed of
intervals of smooth change that are
interrupted by jumps due to vanishing
local minima of the energy ${\cal H}$.
In the intervals between jumps, one can define a {\em linear}
differential polarizability,
$ \chi_{0} \equiv \left<  \eta_{ i} \right>_ i$,
where $\eta_{ i}$ is defined as the {\em linear} response,
$\eta_{ i}=\partial \varphi_{ i}/\partial F$, to a spatially uniform
perturbation
in the drive field.
This linear polarizability is the zero
frequency {\em limit}
of the polarizability, $ \chi_{0} = \chi(\omega  \rightarrow 0)$, since the
jumps are part of the zero frequency response $\chi(\omega = 0)$
\cite{dsf:mft}.
This response is found numerically by taking the derivative
with respect to the drive field $F$
of  Eq.~(\ref{eq:motion}) for a metastable state (where $ \dot{\varphi}_{
i}=0$):
 \begin{equation}
\label{etaeqn}
 \Delta  \eta_{ i} + h \cos( \varphi_{ i}- \beta_{ i}) \eta_{ i} = -1.
 \end{equation}
Given a static configuration $\{ \varphi_{ i}\}$, we determine the $ \eta_{
i}$ by
iterative solution of the diffusion equation \cite{foot1},
 Eq.~(\ref{etaeqn}), by iterating the map
 \begin{equation}
\eta_{ i}  \rightarrow (\sum_{\delta}  \eta_{ i+ \delta} +1)/[2d-h\cos(
\varphi_{ i}- \beta_{ i})],
 \end{equation}
until the fixed point is reached.

\subsubsection{Divergences due to jumps}
\label{sec:divhops}

 Fig.~\ref{micro} is a plot of the linear polarizability $ \chi_{0}$ and
polarization $P$ for the irreversible history
over a small range of reduced field for a single sample of
size $128^{2}$.  This fine field scale allows the individual jumps in the
polarization and corresponding divergences in the linear polarizability to be
clearly seen.
At the point where a metastable configuration vanishes, the polarization
jumps a small amount due to the rearrangement of the phases in some
region.
For the range of fields shown in  Fig.~\ref{micro}, the rearrangement
occurs in regions of a scale of several lattice constants, resulting,
at least well below $ F_{T}$, in a
change of the average polarization $P$
of order $L^{-d} \sim 10^{-4}$.

As the metastable configuration approaches a saddle node
bifurcation at a field $ f_{\rm jump}$,
where the local minima of the energy vanishes, the linear
polarizability from the local degrees of freedom that go unstable
diverges as $( f_{\rm jump} - f)^{-1/2}$.
This divergence leads to the spikes apparent in the linear polarizability
plotted in  Fig.~\ref{micro}.
As the size of the system is increased, these jumps must occur more
frequently.
In the thermodynamic limit, these jumps occur on a set {\em dense}
in
the applied field.
The question arises, then, as to whether the linear polarizability is
well-defined for the typical irreversible approach to threshold.
Surprisingly, the answer is yes.
The contribution to the {\em bulk}
linear polarizability from a single degree of
freedom in a sample of volume $L^{d}$ is \cite{SDF}
 \begin{equation}
\label{deltachi}
\Delta \chi_{0} \sim L^{-d}( f_{\rm jump}-f)^{-1/2}.
 \end{equation}
For a given reduced field $f$, the number of jumps that occur in a small
field interval must be proportional to the volume $L^{d}$ (for $L$ large
enough that the finite size effects discussed below are
unimportant at reduced field $f$).
It follows that the expected distance between jumps (avalanches)
is $\Delta f \sim  n_{\rm av}^{-1} L^{-d}$ for an avalanche
density $ n_{\rm av}(f)$ ($ n_{\rm av}(f)$ is discussed in more detail in
Secs.\ \ref{sec:fss},\ref{sec:discussion} below).
At a given field reduced field $f$,
the probability, $p$, that $\Delta \chi_{0} > \epsilon \chi_{0}$,
for some desired small
relative accuracy $\epsilon$, behaves as
 \begin{equation}
\label{probchihi}
p \sim (\epsilon \chi_{0})^{-2} [ n_{\rm av}(f)]^{-1} L^{-d}.
 \end{equation}
Thus, with probability $(1-p)$
approaching 1 as $L  \rightarrow \infty$, $ \chi_{0}$ is {\em not affected}
by the divergences due to local degrees of freedom going
unstable, to arbitrary accuracy $\epsilon$.
We can therefore examine the nonlinear polarizability, $ \chi^{\uparrow}$,
which
includes the jumps, and the linear polarizability, $ \chi_{0}$,
{\em separately}; both are well-defined.
To study the linear polarizability for
the irreversible history, we examine the {\em median} $ \chi_{0}$, which, for
an
ensemble of a
large number of large systems, will have only a small probability of being
affected by the spikes in the linear polarizability. (In principle, the
above argument implies that the mean $ \chi_{0}$ could have been used, but
the
convergence as $L  \rightarrow \infty$ would be worse.)

\subsubsection{Critical behavior of linear polarizability}

We have calculated the linear polarizability for both the reversible and
irreversible paths in two dimensions.
For the reversible path $\{  \varphi_{ i}^{\downarrow}\}$, we find that
$\lim_{f  \rightarrow 0}\lim_{L  \rightarrow \infty}  \chi_{0}(f)= \chi_{T}$,
where the threshold
polarizability $ \chi_{T}$
is a {\em finite} constant.
The behavior near threshold of $ \chi_{0}(f)$ shows only a power-law cusp.
This is in marked contrast with
finite  systems, where $ \chi_{0}$ diverges as $ \left| f \right|^{-1/2}$.
In order to examine the
leading cusp singularity in $ \chi_{0}$ for the reversible
approach to threshold, it is
better to calculate $d \chi_{0}/dF$, as the extrapolated constant $ \chi_{T}$
is then
unimportant.
We calculate $d \chi_{0}/dF$ directly by solving a linear response
equation similar to  Eq.~(\ref{etaeqn}),
rather than by finding the numerical derivative of $ \chi_{0}$.
We plot the mean of $d \chi_{0}/dF$ in  Fig.~\ref{dchidf}, for
systems of sizes $32^{2}$, $64^{2}$, $128^{2}$, and $256^{2}$.
We define $ \gamma_{\ell}^{R}$ as
the exponent determining the singularity in $ \chi_{0}$  for the reversible
path,
as $f  \rightarrow 0$:
 \begin{equation}
 \chi_{0}(f) =  \chi_{T} - A^{R} \left| f \right|^{- \gamma_{\ell}^{R}}.
 \end{equation}
We find that, for the larger systems, the data for $d \chi_{0}/dF$
is well fit by the form
 \begin{equation}
d \chi_{0}/dF \sim  \left| f \right|^{- \gamma_{\ell}^{R}-1}
 \end{equation}
over more than one decade, with
$ \gamma_{\ell}^{R} =  -0.42 \pm  0.05$.
The coefficient $A^{R}$ is also determined by the fit to the
data for $d \chi_{0}/df$.  Using the fitted quantities $A^{R}$ and $
\gamma_{\ell}^{R}$ for
the reversible path, we estimate
 \begin{equation}
 \chi_{T} \simeq \lim_{f  \rightarrow 0}
\left[  \chi_{0}(f) + A^{R} \left| f \right|^{- \gamma_{\ell}^{R}}\right]
\approx 0.483 \pm 0.005
 \end{equation}
for $d=2$ and $h=5$.

For the irreversible path, the large-volume limit of the derivative of the
linear polarizability,
 \begin{equation}
\lim_{L  \rightarrow \infty} d \chi_{0}(F)/dF,
 \end{equation}
is {\em not}
well-defined, by an argument similar to that given in Eqs.~(\ref{deltachi})
and (\ref{probchihi}).  To determine the behavior of $ \chi_{0}(f)$ as $f
\rightarrow 0$
for this path, we cannot take the derivative, as we did for the reversible
path,
but must examine $ \chi_{0}(f)$ directly.  If a fit is done directly,
allowing
$ \chi_{T}$ to vary, the leading singularity cannot be determined precisely.
Instead, we use the value of $ \chi_{T}$ which has been calculated from the
reversible path.  This analysis assumes that $ \chi_{T} = \lim_{f
\rightarrow 0} \lim_{L
 \rightarrow \infty}  \chi_{0}(f)$ is the same for both paths.
This assumption is consistent with the data;
neither the reversible or irreversible linear polarizability diverges
and they are both increasing, with both paths approaching the unique
threshold configuration \cite{caveat}.
We then fit to the form
$ \chi_{0}(f) =  \chi_{T} -A^{I} \left| f \right|^{- \gamma_{\ell}^{I}}$,
where $A^{I}$ and $ \gamma_{\ell}^{I}$ are
the coefficient and exponent that describe the leading singularity in $
\chi_{0}$
for the irreversible path.
We plot $ \chi_{T} -  \chi_{0}$ for both the reversible and
irreversible path in  Fig.~\ref{chiT-chi}.
The uncertainties in $ \chi_{T}- \chi_{0}$ are larger for the irreversible
path because of the divergences in $ \chi_{0}$ at the jumps.
As mentioned above, we take the median value for
$ \chi_{0}$ on the irreversible path, as the mean would poorly characterize
the
typical value for $ \chi_{0}$.
Our resulting best estimate for $ \gamma_{\ell}^{I}$ is $ -0.40 \pm  0.12$.
The exponents $\gamma_{\ell}^{(R,I)}$ agree to within our error,
though the coefficients $A^{(R,I)}$ are different for the two histories.
This numerical agreement suggests a universality for the singularity in
the linear $ \chi_{0}$, which we now examine in more detail.

\subsection{Linear response: eigenmodes}

We next investigate more generally the linear response about a stationary
configuration.
The linear response
$ \chi_{0}$ can be expressed as the sum of contributions from the
eigenmodes of the operator acting on $\eta$ on the
left-hand side of  Eq.~(\ref{etaeqn}), i.e., the operator
for the linear relaxation of a perturbed
configuration \cite{PBLoned,SNCandDSF}.
For a particular configuration $\{ \varphi_{ i}\}$, we define eigenmodes $
{a^{ {\rm m}}_{ i}}$,
with eigenvalues $-  \Lambda_{ {\rm m}}$, $ {\rm m} = 0, \ldots, N-1$, such
that
 \begin{equation}
\label{eq-eigen}
-  \Lambda_{ {\rm m}}  {a^{ {\rm m}}_{ i}} = \left[  \Delta  {a^ {\rm
m}}\right]_ i + h \sin( \varphi_{ i}-
 \beta_{ i}) {a^{ {\rm m}}_{ i}}.
 \end{equation}
We take the eigenfrequencies to be ordered so that
$0 <  \Lambda_{0} <  \Lambda_{1} <  \Lambda_{2} <
\cdots$.
The polarizability is determined by the relation
 \begin{equation}
\label{eq-eigchi}
 \chi_{0} = L^{-d} \sum_{m=0}^{N-1} (\sum_{ i} {a^{ {\rm m}}_{ i}})^{2}/
\Lambda_{ {\rm m}}.
 \end{equation}
The sum over lattice sites, $\sum_{ i} {a^{ {\rm m}}_{ i}}$, gives the sum of
the
components of the
eigenvectors, which have been normalized so that
$\sum_{ i}( {a^{ {\rm m}}_{ i}})^{2} = 1$.

We have calculated numerically the smallest eigenvalues,
$  \Lambda_{ {\rm m}}$, for $ {\rm m} = 0, \ldots, 5$, in a two-dimensional
system of size
$128^{2}$ for the {\em reversible} extremal history near threshold.
We find that the low lying eigenmodes are {\em localized}: the components
of the eigenmodes are very small outside of a region of size
several times the lattice spacing.
As the threshold is approached, {\em the localization length of the
lowest mode approaches a constant $\approx 1.5$}
(where the localization volume is estimated by the square of
the sum of the components of the normalized eigenmode).
The smallest eigenvalues appear to behave as
 \begin{equation}
 \left|   \Lambda_{ {\rm m}} \right| \sim (f_{ {\rm m}}^{c}-f)^{\mu}
 \end{equation}
for small $f$, with $\mu=0.50 \pm 0.01$, and $f_{ {\rm m}}^{c}$ discussed
below, as shown in  Fig.~\ref{lowmodes}.
Similar results have been found in calculations for the
one-dimensional model with random and incommensurate pinning
phases \cite{PBLoned,SNCandDSF}.
This result is in agreement with a natural picture of the low-lying
eigenmodes consisting of
{\em localized, almost independent degrees of freedom}.  Each eigenmode
approaches a saddle-node bifurcation at a field $ \mbox{$F_{T}^{+}$} +
f_{ {\rm m}}^{c}$, with $f_{ {\rm m}}^{c}>0$ for $m>0$.
In this picture, $\mu=1/2$ exactly, consistent with our numerical results.
At the threshold field, the smallest eigenvalue, $ \Lambda_{0}$, goes to zero
(i.e., $f_{0}^{c}=0$),
with the last minimum of the energy ${\cal H}$ becoming unstable.
The magnitude of the other eigenvalues appear to go to zero at the reduced
fields $f_{ {\rm m}}^{c}$ indicated,
but since the lowest mode has gone unstable
and the configuration is sliding for $f>0$,
the linear analysis clearly no longer applies.

Since the eigenmodes are localized at threshold, the
leading singular behavior can not come from the sum of the
components
of the lowest eigenmode, but rather arises
from the singularities in the eigenvalue distribution for
small $ \Lambda$.
The matrix element $(\sum_{ i} {a^{ {\rm m}}_{ i}})^{2}$ in
 Eq.~(\ref{eq-eigchi}) can be approximated by an average size, $b$, which
approaches a constant as $  \Lambda_{ {\rm m}}  \rightarrow 0$.
The singular part of the linear polarizability is then given
by \cite{PBLoned,SNCandDSF}
 \begin{equation}
\label{eq-chising}
 \chi_{0}^{\rm sing}(f)\sim b \int d \Lambda\,\rho( \Lambda,f)/ \Lambda ,
 \end{equation}
where $\rho( \Lambda,f)$ is the density of eigenvalues at reduced field
$f$.

For the incommensurate model in one dimension,
numerical data suggest that the eigenvalues $  \Lambda_{ {\rm m}}$ obey the
scaling
form \cite{SNCandDSF}
 \begin{equation}
\label{SCDSFscale}
  \Lambda_{ {\rm m}} \sim  \left| f \right|^{\mu}D(m \left| f
\right|^{-\delta}),
 \end{equation}
with $\mu \approx 0.50$  and $\delta \approx 0.18$;  for incommensurate
pinning, the scaling function $D$ is not continuous and is invariant only
under discrete rescalings.
The exponent $\mu$ is interpreted as determining the frequency of the
softest modes (``active regions'') near the depinning transition.   The
exponent $\delta$ describes the scaling of the density of these regions.
These exponents yield the dependence on frequency $\omega$ of the
ac conductivity $\sigma(\omega)$ at threshold to be
$\sigma(\omega) \sim \omega^{\delta/\mu}$, consistent with numerical
results for the incommensurate model \cite{SNCandDSF}.

For a continuous distribution of eigenvalues $\rho( \Lambda,f)$, this scaling
form can be rewritten as a scaling form for the density of states
$\rho( \Lambda,f)=dm/d \Lambda$ by solving  Eq.~(\ref{SCDSFscale}) for $m$
and
differentiating.  The result is
 \begin{equation}
\label{eq-dos-scaling}
\rho( \Lambda,f) \sim
 \Lambda^{\alpha} \hat{\rho}( \Lambda \left| f \right|^{-\mu}),
 \end{equation}
where $\alpha=(\delta-\mu)/\mu$ and the scaling function $ \hat{\rho}$
approaches a constant as its argument becomes large.
The exponent $\alpha$ defines the distribution of the modes at threshold
($f=0$) for small eigenvalue $ \Lambda$.
The exponent $\mu$ characterizes the frequency scale at which
the distribution of modes for $f<0$ differs significantly from the threshold
distribution.  This is a plausible form for the form of the density of
states, especially for the reversible path.  If the field is lowered from its
threshold value by a small amount, the configuration will change very little
and the only modes which will have significantly different eigenvalues will
be those modes which are the softest at threshold.  This scaling form is
certainly consistent with our results for the behavior of the individual
eigenmodes as $f  \rightarrow 0^{-}$ for the reversible path in two
dimensions.
For this reversible path, we thus expect $ \hat{\rho}(u)=0$ for $u$ less than
a value $u_{c}$, indicating the absence of modes with frequencies less
than $u_{c} \left| f \right|^{\mu}$.
We now investigate the consequences of assuming this scaling {\em form}
for the density of states for {\em both} histories, albeit with different
scaling {\em functions}.

The scaling relation  Eq.~(\ref{eq-dos-scaling}), taken together with
 Eq.~(\ref{eq-chising}), determine the form of the linear polarizability for
small
$f$ on the reversible path.
{}From  Eq.~(\ref{eq-dos-scaling}),
the singular part of the linear polarizability has the form
 \begin{equation}
 \chi_{0}^{\rm sing}(f)\sim f^{\mu\alpha},
 \end{equation}
implying the scaling relation
 \begin{equation}
\label{eq-gamma-scaling}
 \gamma_{\ell} = -\mu\alpha.
 \end{equation}
In two dimensions, this scaling relation implies, given our computed values
of $\mu$ and $ \gamma_{\ell}$, that $\alpha =  0.84 \pm  0.12$.
Since $\alpha$ is defined by the distribution of states at threshold,
which is independent of history, the numerical agreement of the
exponents $ \gamma_{\ell}^{I}$ and $ \gamma_{\ell}^{R}$ suggests that the
exponent $\mu$
is the same for the two histories; it is presumably exactly $1/2$.
Of course, the scaling function $ \hat{\rho}$ is different for the two
histories.

 Fig.~\ref{dospict} schematically shows the density of states for the two
paths that we have examined.
{}From the observation that $  \Lambda_{ {\rm m}} \sim (f_{ {\rm
m}}^{c}-f)^{\mu}$ for
small
$f$ on the {\em reversible path}, it can be seen that there is a gap of size
$f^{\mu}$ in the density of states for configurations on the reversible path.
In  Ref.~\cite{dsf:mft}, it is argued that, for configurations along the
{\em irreversible  path},
a density of states that is {\em linear at small}
$ \Lambda$ is stable to changes in the field.
These considerations lead us to speculate
that the density of states for the irreversible and
reversible approaches to threshold are as
shown in  Fig.~\ref{dospict}.
We conclude that although the two histories have very different
densities of
states at the same field, due to the difference in their scaling
functions $ \hat{\rho}$, there is a common underlying frequency scale
defined by $f^{\mu}$, which, with the distribution of eigenmodes at
threshold, characterized by the exponent $\alpha$,
determines the singularity in the linear response.

\section{Finite-size effects and avalanches}
\label{sec:fss}

In the study of conventional thermodynamic transitions, the concept of a
dominant diverging length scale, the correlation length, plays a crucial
role in the understanding of scaling relations and the physical
description of the system.
Likewise, correlation lengths for some deterministic dynamical
systems \cite{Liap,PAetal} have provided insight into the
behavior of those models.
In order to better understand the nature of the CDW depinning transition, it
is important to develop an understanding of the characteristic length(s).
For CDW's, there have been some attempts to understand the
correlation length in the sliding state
numerically \cite{miscnumerics,SibaniPBL,Matsukawa} and there has
been some success very recently in expanding about mean-field theory
in $d=4-\epsilon$ dimensions \cite{narayanfisher}.
These results do not, however,
address the static behavior in the pinned phase.

In this section, we address the question of the definition and behavior of
correlation lengths in the static regime for the lattice CDW model with
random pinning phases.
We define a finite-size-scaling length using the
distribution of threshold fields and determine
the corresponding finite-size-scaling
exponent $ \nu_{\rm T}$, which we find to be very close to $2/d$ in both one-
and
two- dimensional systems.
We also give numerical results on finite-size scaling of the
polarization for the irreversible path;
from these, we determine an exponent $ \nu_{\rm n}$
which describes the finite-size crossover for the polarization which is,
within numerical accuracy, equal to the value for $ \nu_{\rm T}$.
The sizes of ``avalanches'', which occur when a local mode becomes unstable,
are found to have a maximum typical size that diverges in a fashion
consistent with that given by $ \nu_{\rm T}$ and $ \nu_{\rm n}$.
By contrast, the finite-size crossover
for the {\em reversible} polarizability scales very differently in two
dimensions, with an exponent $ \nu_{\ell}=0.44\pm 0.05$.  We discuss these
relationship of these results and their connection with the bound $ \nu_{\rm
f}\geq
2/d$, for
finite-size-scaling exponents $ \nu_{\rm f}$, proved by Chayes, {\em et
al} \cite{chayesetal}.

\subsection{Finite-size scaling}

We first briefly review the theory of finite-size scaling, which has
been very useful for the numerical study of conventional
critical points \cite{FSS}.
Suppose that in an infinite system, some
quantity $Y$ scales as $Y \sim \delta^{-y}$, where
$\delta$ is the
reduced control parameter, which goes to zero at the critical point, and the
exponent $y$ describes the critical behavior for $Y$.  In a {\em finite}
system of linear size $L$ much larger than any microscopic length, the
finite-size scaling hypothesis states that
 \begin{equation}
Y(\delta,L) =  \left| \delta \right|^{-y}\Psi(L\xi^{-1}),
 \end{equation}
where $\xi \sim  \left| \delta \right|^{-\nu}$ is the correlation length and
$\Psi$ a universal scaling function which can, however, depend on
the type of boundary conditions.
We consider only periodic boundary conditions, which are the simplest.
Given data on the
behavior of $Y(\delta,L)$, the exponents $\nu$ and $y$ can be extracted by
finding values of these exponents for which a scaling plot of
$Y \left| \delta \right|^{y}$ vs.\
$L \left| \delta \right|^{\nu}$ yields (asymptotically) a single curve.
In this section, we apply such a
finite-size-scaling analysis to the study of static properties of CDW's
as $ F_{T}$ is approached from below.

For disordered systems,  a finite-size-scaling length can be defined in
terms of the
statistical properties of a large number of finite-size
samples \cite{chayesetal}.  Such a length can be
defined by the behavior of the
probability of a finite-size-scaling event; the occurrence of such an event
in each sample depends on the realization of the disorder for
that sample and the value of the disorder parameter.
For example, in a bond-percolation model a finite-size-scaling
event may be defined as
the existence of a path of bonds that connects two sides of a finite-sized
system:
in large samples
the probability of such a path existing varies rapidly with
the bond probability $p$ near the
percolation threshold $p_{c}$.
The scaling of the
probability distribution for this event defines a finite-size length scale.
In  Ref.~\cite{chayesetal}, it is proven that, when the transition occurs
at a non-trivial value of the disorder parameter,
the exponent $ \nu_{\rm f}$ for the divergence
of such a finite size scaling length must satisfy the bound
$ \nu_{\rm f} \geq 2/d$.

It is important to make a distinction between the {\em statistical behavior}
of the model as a function of the different
realizations of the disorder and the behavior of {\em a single sample}.
The bound of Ref.~\cite{chayesetal} applies to the statistical
behavior {\em only} and does not necessarily apply to
the finite-size effects in a single sample, which may be very different.
For example, consider the simple case of an Ising magnet at low temperatures
in $d \ge 3$ which, in an
infinite-volume sample,
undergoes a first order transition as the magnetic field $H$ passes through
zero.  At this transition, the mean
magnetization density $m$ jumps between the values $m_{0}$ and $-m_{0}$.
We now introduce an independent random magnetic field
$ h_{ i}$, at each site.
In a given finite sample, an approximate transition field can be defined
as the value, $H_{c}(\{ h_{ i}\})$, of the uniform field $H$ for which
the thermal expectation of the magnetization equals zero.
For a collection of finite-size samples, of
size $L^{d}$, the width of the distribution over the random fields
of the approximate transition fields $H_{c}(\{ h_{ i}\})$
has width $L^{-d/2}$;  this can be seen
by noting that the sample-to-sample variations in the spatially averaged
random field of the samples will have
variations of this magnitude.
This implies a finite-size-scaling exponent $ \nu_{\rm f}=2/d$
\cite{Imbrie}.

The width of the transition in a {\em single} sample, on the other
hand, can be defined
as the range of applied magnetic fields over which the
magnetization switches between some $m_s$ and $-m_s$, for a given $m_s$,
for example the range $H(m=m_s)-H(m=-m_s)$), with $m_s=m_0/2$.
The width of this transition {\em in a single sample}
scales as $L^{-d}$: for a range of fields
$\delta H\approx TL^{-d}$ (T being temperature), the difference in free
energy between the $+m_{0}$ and $-m_{0}$ configurations will be $O(T)$,
so that the thermal average $m$ will have a magnitude considerably less
than $m_{0}$ for fields within this
distance of the transition $H_{c}(\{ h_{ i}\})$
{\em of the single sample.}
 In contrast
with the finite-size effects for the {\em statistical} behavior, this
suggests a finite-size-scaling exponent $\nu=1/d$ for individual
samples.

\subsection{Threshold field distribution}

The threshold field is the first quantity that is calculated for each
realization in our numerical study of CDW's, and it is a
natural quantity to study for finite-size effects.
We examine the probability
distribution of the threshold field for randomly chosen pinning phases
$\{ \beta_{ i}\}$.
For an infinite system, there should be a single value for the
threshold field,
$ F_{T}(\infty)$, so that as $L \rightarrow\infty$,
the probability distribution of the threshold field should
approach $\delta( F_{T}- F_{T}(\infty))$.
In finite size systems, however,
due to the variations in the pinning from sample to
sample, the threshold field probability distribution will
have a finite width.

The threshold field averaged over many realizations of the pinning for
systems of linear size $L$, $ \overline{ F_{T}}(L)$, is plotted in
Fig.~\ref{Ft}.
The average threshold field rapidly approaches a constant as
$L  \rightarrow \infty$.
It is difficult to study finite size effects in
the mean threshold field, as the
difference $ \overline{ F_{T}}(L)- F_{T}(\infty)$ is smaller than the
statistical error for the
largest samples.
Here we focus on the second moment of the distribution, noting the
possibility
that $ \overline{ F_{T}}(L)- F_{T}(\infty)$ might scale {\em differently}.

In  Fig.~\ref{typFtdist}, we plot a histogram distribution
for the computed values of the threshold field for $128$ sample
systems of size $32^{2}$.
{}From such a sample distribution, for systems of various sizes $L$, we can
determine an estimate for the widths of the distribution at various scales.
We characterize such a distribution
by computing its r.m.s.\ width $ \mbox{$\Delta F_{T}(L)$}$.

This quantity, $ \mbox{$\Delta F_{T}(L)$}$, is directly related to a
finite-size-scaling
length, in the sense of  Ref.~\cite{chayesetal}.  We choose the finite-size
scaling event to be the sliding of the CDW.  The width of threshold fields
$ \mbox{$\Delta F_{T}(L)$}$ then defines a field scale over which the
probability of this
finite-size-scaling event changes significantly.  The occurrence of this
event
depends on the randomly chosen pinning phases and the value of the control
parameter.  The parameter that controls the disorder is the pinning strength,
$h$.  If the infinite-system threshold field has a dependence on
$h$, $ F_{T}(h)$, that is well behaved, with $d F_{T}/dh \neq 0$,
$F$ can be taken as an equivalent parameter.
Since we expect that $ F_{T}(h)$ is a smooth
function, the finite-size-scaling exponent that we derive from
$ \mbox{$\Delta F_{T}(L)$}$ should satisfy the bound of
Ref.~\cite{chayesetal}.

In  Fig.~\ref{DFt} we plot our results for the width of the threshold field
distribution $ \mbox{$\Delta F_{T}(L)$}$ vs.\ linear dimension $L$ for one-\
and two-
dimensional CDW's.
The lines show the least-square fits to the form
 \begin{equation}\label{nutdef}
 \mbox{$\Delta F_{T}(L)$} \sim L^{-1/ \nu_{\rm T}},
 \end{equation}
where $ \nu_{\rm T}$ is a finite-size-scaling
exponent for the transition between the static and sliding states.
To ensure that the length scale that we measure is much greater than any
microscopic length scale ($ \xi_{o}$ or the lattice spacing), we fit to
systems of
size $L\ge 16$ (the fit would be much worse if the $L=8$ data were included).
{}From these fits, we derive a value for $ \nu_{\rm T}$ of $ 2.01\pm 0.02$
in one dimension and $ 1.01\pm 0.03$ in two dimensions.
Within our statistical error,
these results satisfy, and appear to saturate, the bound $ \nu_{\rm f} \geq
2/d$ of
 Ref.~\cite{chayesetal}.
In the infinite range model, the width of the distribution of threshold
fields
as a function of number of degrees of freedom obeys
$\Delta  F_{T} (N) \sim N^{-1/2}$ \cite{uniqueaam}.
If one naively extends this
result to short-range interactions, with $N=L^{d}$, one would
obtain $ \nu_{\rm T} = 2/d$.
For the ``random friction'' model \cite{Mihalyetal},
the threshold field can be calculated
explicitly \cite{Webman87,snc}, and
the exponent $ \nu_{\rm T}$ is exactly $2/d$, since the threshold field is
just the
average of the pinning strengths.
It is not at all clear how to show that a similar result should
hold for the finite-dimensional CDW model; indeed one might expect a
non-trivial exponent, at least in low dimensions, and it is quite possible
that $ \nu_{\rm T}=2/d$ only represents an approximate value.

Recent renormalization group calculations by Narayan and Fisher
\cite{narayanfisher} have found that the exponent for the distribution
of threshold fields is equal to $2/d$ to lowest order in $d=4-\epsilon$,
and there is some indication that this may be true to all orders in
$\epsilon$, but perhaps with non-perturbative corrections.
An interesting open question is whether the distributions of
threshold fields for large systems is Gaussian; this is probably
related to the question of whether $ \nu_{\rm T} = 2/d$.

\subsection{Finite-size effects in polarization and polarizability}

Examining finite-size effects of quantities other than the threshold field
gives us a check on the results for $ \nu_{\rm T}$
and allows us to investigate the possibility of the existence of more than
one important length scale.

One such quantity is the polarization, $P$, for the {\em irreversible}
history (i.e., $\{  \varphi_{ i}^{\uparrow}(F)\}$ near $
\mbox{$F_{T}^{+}$}$).
As discussed earlier, the polarization in the stationary phase is bounded in
a
finite system.
The critical divergence in the polarization of the infinite system must
therefore be cutoff at some field scale, which depends on the size of the
system.  In  Fig.~\ref{pol_one} and  Fig.~\ref{pol_two} we plot the
polarization for
one- and two-dimensional systems of various sizes.
Apparent in these plots  is a crossover from the divergent large-system
critical behavior to a finite value of $P$,
which is plotted in  Fig.~\ref{threshPvsize}.

We first examine the finite-size scaling in the case of
{\em two dimensions}.
We assume a scaling form for the mean polarization:
 \begin{equation}
\label{eq:pscale}
\overline{P}=f^{-\gamma+1}\hat{P}(Lf^{ \nu_{\rm n}}),
 \end{equation}
with $\hat{P}$ approaching a constant for large values of its argument and
behaving as $\sim x^{(\gamma-1)/ \nu_{\rm n}}$ for small arguments $x$,
consistent with the observation of constant polarization at small $ \left| f
\right|$.
 Fig.~\ref{polscl2} shows a scaling plot of $\overline{P} \left| f
\right|^{\gamma-1}$
vs.\ $L \left| f \right|^{ \nu_{\rm n}}$, for fitted exponents $ \nu_{\rm n}=
1.0\pm 0.1$
and $\gamma=1.8\pm0.1$ (where the errors are
estimated by finding what values of the
exponents give an unacceptable deviation from a single curve).
These values are in agreement with our earlier estimate for $\gamma$,
based on the nonlinear polarizability $ \chi^{\uparrow}$, and are also
consistent with
the numerical equality of $ \nu_{\rm T}$ and $ \nu_{\rm n}$.
It follows directly from the assumed behavior for the scaling form that
 \begin{equation}\label{Pscl:rel}
 \nu_{\rm n}=(\gamma-1)/\rho,
 \end{equation}
where the polarization at threshold $\overline{P_{\rm T}} \sim L^{\rho}$,
as shown in  Fig.~\ref{threshPvsize}.
The scaling relation  Eq.~(\ref{Pscl:rel}) is found to be
satisfied by our exponents for two-dimensional systems.
Note however that, for $f<L^{1/ \nu_{\rm n}}$, the sample to sample
variations
of the polarization are of the same order of magnitude as the mean
polarization.

It is not possible to find a good fit to a
single scaling form for the polarization
data in one dimension.
We instead estimate $ \nu_{\rm n}$ by a cruder
procedure, using the observation that the polarization approaches a constant
at small $ \left| f \right|$.
We can define a crossover reduced field, $ f_{X}(L)$, by
$P_{T} = P_{0} ( f_{X})^{-\gamma+1}$, with $\gamma$ and $P_{0}$ determined
from the divergent behavior of $ \chi^{\uparrow}$ in the largest systems.
This is
consistent with assuming the scaling form of  Eq.~(\ref{eq:pscale}).
{}From the data of  Fig.~\ref{threshPvsize}, we thereby
derive an exponent $ \nu_{\rm n}$ for the
finite-size effects in the polarization, using $ f_{X} \sim L^{-1/ \nu_{\rm
n}}$.
Using this relation, we find that $ \nu_{\rm n}= 2.0\pm 0.5$,
in numerical agreement for our value for $ \nu_{\rm T}$ in one dimension.

In  Fig.~\ref{polscl1} we plot the one-dimensional data on a scaling plot,
using
the above {\em calculated} value for $ \nu_{\rm n}$.
This plot suggests a consistent explanation
for the failure to find a single fit
for the data of various sizes: we are not close enough to threshold to see
the asymptotic scaling behavior.
The larger systems appear to be approaching a single
scaling form, for low fields ($ \left| f \right| < 0.1$), while the smaller
systems seem
to deviate strongly.  Our large value for $ \nu_{\rm n}$ implies that very
large
systems must be examined to see clearly the true scaling behavior.

\subsection{Relation of finite-size effects}

Our results for the finite-size effects of the irreversible polarization and
comparison with the results for $ \nu_{\rm T}$ suggest a picture for a
diverging
correlation length near the depinning transition characterized by an
exponent $\nu= \nu_{\rm T}= \nu_{\rm n}$.
As the threshold field is approached in an infinite system along the
irreversible path, regions of size $ \left| f \right|^{-\nu}$ are subject to
fields that
exceed the threshold field for these subsystems, if they were to be
considered
independently.
These regions slide forward some distance, but are prevented from sliding
further by the subsystems where the threshold field has not
been exceeded;  on scales large than $ \left| f \right|^{- \nu_{\rm T}}$, the
probability of the
subsystem threshold being exceeded is small.  There are, therefore, a series
of ``avalanches'' which occur on length scales up to $\xi\sim \left| f
\right|^{- \nu_{\rm T}}$,
as the threshold field is approached.
The jumps in the polarization thus grow as the size of the avalanches grow.
The polarizability and polarization on the irreversible
path can thus diverge as threshold is approached in an infinite system.

In a finite system, if an avalanche of size of order the system
size occurs, the whole system will start sliding.
The average difference of the reduced fields at which thresholds of
subsystems of size $L$ occur is of the order of $L^{-1/ \nu_{\rm T}}$.
For reduced fields of order $L^{-1/ \nu_{\rm T}}$, there will thus be a
leveling off
of the polarizability, as there will be no avalanches of linear size of order
$L$ in this range, but subsystems smaller than the system size will continue
to depin at the same rate, as the width of the distribution of the
threshold field on these length scales is much larger than $L^{-1/ \nu_{\rm
T}}$.
This physical picture is supported by the data of  Fig.~\ref{chi-two}, where
the
{\em polarizability} appears to roughly level off at a value that
increases with system size.
With the polarizability approaching a constant, the polarization saturates.
This is consistent with a crossover in the polarization curve at a reduced
field scale given by the exponent for the threshold field distribution, $
\nu_{\rm T}$.
We therefore conjecture that
 \begin{equation}
 \nu_{\rm n}= \nu_{\rm T}\equiv\nu.
 \end{equation}
We emphasize that the definition of the
finite-size-scaling length for the polarization is {\em not} based on the
scaling of the probability distribution for some finite-size-scaling event,
and it is therefore necessary to argue, as we have here, that
the exponents $ \nu_{\rm T}$ and $ \nu_{\rm n}$ are directly related, since
we
cannot prove $ \nu_{\rm n} \ge 2/d$ directly.

\subsection{Avalanches}
\label{subsec:avalanches}

The above argument relating avalanche size to the definition of $ \nu_{\rm
T}$ can be
compared with the numerically calculated distribution of avalanche events.
We have conducted such a calculation for a system of size $256^{2}$.
Starting from the negative threshold field $ \mbox{$F_{T}^{-}$}$, we
adiabatically
increase the field $F$, thereby following the path $\{  \varphi_{
i}^{\uparrow}(F)\}$ in
configuration space.
At each local instability,
we measure the moment $ \Delta P$, defined as the change in polarization
$P$ from just below to just above the instability.
The no-passing rule can be used to show that, for an infinitesimal change in
the field, no phase may advance more than $2\pi$.
The change in phase $\Delta \varphi_{ i}$ at each site during an avalanche is
therefore bounded by $2\pi$, and the quantity $ \Delta P/2\pi$ provides
a good estimate of the avalanche size, as the width of the ``boundary'' of
the avalanche is of the order of the Lee-Rice length (qualitatively,
the avalanches appear to be compact and not fractal in our simulations).

The results of this calculation are plotted in  Fig.~\ref{avalanche}.
Each point represents a single avalanche (event) due to a local instability
(corresponding to the peaks in $ \chi_{0}$ and discontinuities in $P$ shown
in  Fig.~\ref{micro}), plotted in the $ \Delta P$-$F$ plane.
According to the arguments in the previous section, there should be a scale
$\xi$ which determines avalanche sizes and diverges as
$\xi\sim(F- F_{T})^{-\nu}$.
The solid line in  Fig.~\ref{avalanche} shows the expected dependence
of avalanche size on field, $ \Delta P\sim(F- F_{T})^{-d\nu}$, under
the assumption that $\nu= \nu_{\rm T}$ (we have added an arbitrary vertical
shift
in the curve).
The dependence of the apparent cutoff in the measured avalanche sizes on
the field is consistent with this dependence.
A more detailed analysis of the cutoff length
would require a greater number of events
than can be feasibly obtained at this time.

To investigate the limit $F\nearrow \mbox{$F_{T}^{+}$}$, we find the
cumulative
distribution of the events contained in the region outlined by the dashed
lines shown in  Fig.~\ref{avalanche}.
For this set of avalanches, the cutoff length is greater than the
linear size of the avalanches, and therefore these data should reflect the
distribution of avalanche sizes just below
the threshold field in an infinite system (we have tried several
different reasonable criteria for choosing sets of avalanches near
threshold, and find results independent of the exact choice).
A logarithmically binned distribution of these events is shown
in  Fig.~\ref{avdist}, where $N(\Delta P)$ is the number of events
in the range $[2^{-1/2}\Delta P,2^{1/2}\Delta P]$.
To within the statistical errors, this distribution is fit by the
power law form
 \begin{equation}\label{kappadef}
N( \Delta P)\sim  \Delta P^{-\kappa/d}
 \end{equation}
with $\kappa= 0.34\pm 0.10$.
The value of this exponent is in numerical agreement with that of the
corresponding distribution seen in models of self-organized criticality
\cite{marg-stable,betacorrection}.
This strongly suggests that the state on approach to threshold in CDW's is
closely related to the self-organized critical state seen in
``sandpile'' models.
This is consistent with a distribution of avalanches cutoff only
by the system size in the range of reduced fields $L^{-1/ \nu_{\rm T}}<f<0$
in the CDW models and a distribution of responses to
perturbations similar to that seen in the sandpile models.
These results suggest a possible
universal behavior for the nonlinear response in systems in
such critical states, whether obtained by adjusting a control
parameter near threshold or by a mechanism of ``self-organization''.
Note that the {\em typical} avalanche has a {\em non-divergent}
size as $F\nearrow F_{T}$,
since $\kappa>0$ --- the diverging correlation length becomes
evident only in the tail of the avalanche size distribution.
The {\em average} avalanche size does diverge as threshold is approached,
accounting for the diverging polarizability on the irreversible
path (see  Sec.~\ref{sec:discussion} below).

\subsection{Effects of finite-size for the reversible path for $d=2$}

We can also define a finite-size-scaling exponent,
$ \nu_{\ell}$, for the crossover in
the linear polarizability, for the {\em reversible} history.
Since our value for $ \chi_{T}$ in the limit of an infinite-size system is an
extrapolation which can only introduce error into the study of finite size
effects, we examine, as above, the derivative
of the linear polarizability, $d \chi_{0}/df$, for
finite-size effects.
In  Fig.~\ref{dchidf}, we have plotted $d \chi_{0}/dF$ for several system
sizes
in two dimensions, averaged over several realizations for each size.
In  Fig.~\ref{dchidfscl}, we show scaled data $(d \chi_{0}/dF) \left| f
\right|^{ \gamma_{\ell}^{R}+1}$
as a function of $ \left| f \right|L^{1/ \nu_{\ell}}$, to test the scaling
form
 \begin{equation}
\label{eq-chi-scaling}
\frac{d \chi_{0}}{dF} =  \left| f \right|^{- \gamma_{\ell}^{R}-1} X(L \left|
f \right|^{ \nu_{\ell}}),
 \end{equation}
with $X(z)  \rightarrow \infty$ as $z  \rightarrow \infty$, and $
\gamma_{\ell}^{R}$ as determined
earlier.
The value of $ \nu_{\ell}$ which gives the curves shown in
Fig.~\ref{dchidfscl}
is $ \nu_{\ell} =  0.44 \pm  0.08$.

Our numerical result for the value of $ \nu_{\ell}$, which is {\em less} than
$2/d$,
implies that the natural definition for the finite-size length scale on the
{\em
reversible} path, does not satisfy the definition of a finite-size length
scale in
the sense of  Ref.~\cite{chayesetal}.
The distinction between the finite-size-length scale for the linear
polarizability and the behavior of sample-to-sample fluctuations can be
understood by examining the effects of finite-size on the linear
polarizability for the reversible path.

On the reversible path, there are no hops, and no triggerings of any
``avalanches''.
The configuration at reduced field $ \left| f \right| \ll 1$ is connected to
the
threshold configuration by a continuous path.  The scaling of the
threshold avalanche distribution thus does not enter.  The polarizability of
configurations on this path in the finite-size case can be treated as in
Sect.~(3.3.3), with an integral over a spectrum replaced by a sum over
discrete values.  It is easily seen that the polarizability of the
softest mode dominates the finite-size effects.  The contribution of the
softest mode to the bulk polarizability $ \chi_{0}$ is
 \begin{equation}
\Delta \chi_{0}\sim L^{-d} \left| f \right|^{-\mu},
 \end{equation}
since the softest mode goes unstable at $ \left| f \right|=0$.
When $\Delta \chi_{0}$ becomes of
the same order as the singular part of $ \chi_{0}$, there will be a crossover
from the bulk critical behavior to {\em the single particle behavior}.
This crossover will occur at reduced field
 \begin{equation}
 \left| f \right| \sim L^{d/( \gamma_{\ell}-\mu)}.
 \end{equation}
This implies that the finite-size-scaling exponent for the linear
polarizability on the reversible path satisfies the scaling
relationship
 \begin{equation}\label{nulscl}
 \nu_{\ell}=\frac{1}{d}(\mu- \gamma_{\ell}).
 \end{equation}
The relationship  Eq.~(\ref{nulscl}) is consistent with our numerical values
for these exponents in two dimensions.
The finite-size effects on the reversible
path are due to the behavior of the softest mode in each sample
near to {\em its} threshold and not the
probability for some finite-size-scaling event to occur, which would
depend more strongly on the
realization of the disorder.

\section{Discussion}
\label{sec:discussion}

In this final section, we compare our results with those on related
models and, qualitatively, with experiments on CDW's below threshold, as
well as raising general questions about universality and scaling.  We
first recapitulate our main results.

On the irreversible approach to threshold (starting from the
negative threshold),
the motion of the CDW consists of smooth motion superimposed on a series
of jumps or avalanches that result
from the local minimum of the energy in which
the system lies becoming unstable.
On increasing $F$, the polarizability
$ \chi^{\uparrow}$ diverges with an exponent $\gamma$ dominated by the jumps.
The avalanches are initiated by {\em local} instabilities via
simple saddle-node bifurcations (i.e., the vanishing of the eigenvalue
for a localized mode), though the avalanches can be quite
large, with an apparently power law distribution of their sizes
extending out to the correlation length.
The low frequency {\em linear} response in contrast
is dominated by the local modes which are nearly unstable.
The divergences of $ \chi_{0}$ at each of the local
instabilities typically contributes a negligible amount in the large
system limit, so that $ \chi_{0}(F)$ is a well-defined smooth function in
this
limit.
In an infinite system,
the zero-frequency linear response $ \chi_{0}$ approaches a {\em finite}
value
$ \chi_{T}$ at threshold with an upwards cusp characterized by an exponent
$- \gamma_{\ell}>0$; this is in contrast to the divergence of $ \chi_{0}$ in
a finite
system at threshold.

Associated with the non-linear behavior of the avalanches, we have found
evidence for a characteristic length, $\xi\sim( F_{T}-F)^{-\nu}$ with
$\nu\approx 2/d$, appearing in the width of the distribution of
threshold fields in finite-size systems or the finite-size rounding of
the polarization divergence.
In two dimensions,
this exponent is also consistent with rough estimates of the divergence of
the maximum likely avalanche size as threshold is approached
on the irreversible extremal path.
We believe therefore that this length is the
correlation length associated with the dominant physics.
The exponent $\nu$ should satisfy the bound $\nu
\ge 2/d$ and appears to be close to saturating this bound in the
dimensions ($d=1,2$) that we have studied.

We believe that the irreversible approach to threshold described above is
generic in the sense that the same behavior will be found on approaching
threshold from all except specially prepared initial conditions, although
the {\em amplitudes} of the divergences in $ \chi^{\uparrow}$ and $\xi$ may
differ.
Experimentally, the polarization on the generic approach to threshold
does not exceed that given by an average CDW displacement of more than
several CDW wavelengths \cite{NMR,Ong:pol}
(a polarization by one CDW wavelength is
$P=2\pi$ in the dimensionless units here).
This is consistent with
our results for three-dimensional CDW's, shown in  Fig.~\ref{polthreed}.

The behavior of specially prepared initial conditions can be
strikingly different from the generic behavior.
When the field is reduced from threshold, there
appears to be a region of finite width in $F$ over which there are no
jumps and the polarization is fully reversible, so that
$ \chi^{\uparrow}= \chi^{\downarrow}= \chi_{0}$.
It appears quite likely that this will persist even in infinite systems,
especially if the distribution of pinning strengths is bounded away from
zero.
As noted earlier, the qualitative features of the reversibility
found in our numerical simulations is consistent with
experiment \cite{Ong-hyst}.
Although we have not been able to find a type of rare region which would
invalidate this conclusion, and our numerical evidence does not
appear to indicate
even logarithmic dependence on size of the lower limit of reversibility
$F_{R}^{+}$, this conclusion should nevertheless be regarded with some
caution.

In any case, the behavior of the polarizability in the reversible
approach back to threshold (after $F$ is lowered to a field above
$F_{R}^{+}$), is characterized by only an upwards cusp in $ \chi_{0}$ to a
constant, $ \chi_{T}$, at threshold.
To numerical accuracy, both the {\em value}, $ \chi_{T}$, at threshold in an
infinite system, and the {\em exponent} $- \gamma_{\ell}$ characterizing the
cusp, are the {\em same} as found for the linear polarizability $ \chi_{0}$
in
the irreversible approach although the {\em amplitudes} of the cusps differ.
Thus there is evidence for some degree of universality in the linear
response in spite of the radical difference of the non-linear behavior in
the two approaches to threshold.

This can be interpreted in terms of scaling functions for the
distribution of relaxational frequencies of the linear modes
which are characterized
by history independent {\em exponents} ($\alpha$ and an apparently trivial
exponent $\mu \approx 1/2$) but history dependence of the actual scaling {\em
functions}.
The reason for this behavior is somewhat of a mystery, however, because
of the rather large rearrangement of the distribution of modes following
avalanches as is discussed below.
The finite size corrections to the linear polarizability in the
reversible regime of {\em each particular sample} are characterized by a
length which diverges more slowly than $\xi$ with an exponent $ \nu_{\ell}
<2/d$.
It appears that this is {\em not} a true correlation length because it
essentially is determined by the crossover from behavior of the large
system to that of a single {\em localized} mode, rather than to a
collective property of the whole system; it is a length
related to a power of the
density $(\rho^{-1/d})$ of soft modes, rather than a length related
to a response on large scales.
Because of the existence of a sharp threshold in finite systems and the
absence of a smooth connection between the pinned and moving phases, many
quantities which naively appear to be characteristic lengths can occur.
Similar behavior is found in other types of collective non-linear transport
with a sharp threshold \cite{rivers}.

Before further analysis of the implications of our results, which are
summarized in the Table, we first
briefly compare them with those on related models.

\subsection{Comparison with related models}

In mean field theory --- valid for infinite range stiffness of the CDW
--- the critical behavior of the polarization
depends, as in short-range systems,
on the approach to threshold, but, in addition, it
depends on the distribution of the pinning strengths \cite{dsf:mft}.
Consider the generic case of a distribution of pinning strengths, $\{ h_{
i}\}$,
which is bounded below by a value $h_{0}$.
If $h_{0}$ is not too small, then the following behavior is found: on the
first approach to threshold, the polarizability $ \chi^{\uparrow}$ diverges
with an
exponent $\gamma>0$ which depend on the form of the distribution of the
$\{ h_{ i}\}$.
As the field is decreased from threshold, there is a regime of reversible
behavior \cite{foot:smallh} characterized by a polarizability $ \chi_{0}$
which goes to a constant at threshold with an upwards cusp with
a non-universal exponent $ \gamma_{\ell} < 0$.
If the field is decreased to $- F_{T}$ and then increased again, the
polarizability will again diverge.
Thus the behavior we have found in this paper is quite similar
qualitatively to that of mean field theory, although we expect the latter
to be less universal (see below).

The one-dimensional incommensurate CDW model has also been studied quite
extensively \cite{SNCandDSF}.
Although there are again many possible approaches to
threshold, the only one which has been studied is the approach from the
$F=0$ ground state, which is reversible all the way from $- F_{T}$ to $
F_{T}$,
a consequence of the identical pinning strengths and special symmetry.
As the threshold is approached, the polarizability {\em diverges} with
$ \gamma_{\ell} \approx 0.34$.
This is associated with a distribution of local relaxational frequencies
the lowest of which vanishes at thresholds with an exponent $\mu=0.50
\pm 0.005$, consistent, as are our data, with $\mu=1/2$.
Near threshold, the distribution of these frequencies has the form
 Eq.~(\ref{eq-dos-scaling}) with $\alpha\approx -0.68$.
A characteristic length diverging with exponent $ \nu_{\ell}=\mu-
\gamma_{\ell}$ is also
found. The primary difference between this behavior and the behavior of
the random 2-$d$ system studied here on its reversible approach, is
associated with the sign of $ \gamma_{\ell}$.
In both cases, $d \nu_{\ell}$ should be interpreted as the density of spatial
modes with frequencies of order $ \left| f \right|^{\mu}$.
[For the 1-$d$ incommensurate model, there is of course no distribution
of threshold fields, but the size dependence of the threshold field
itself converges very rapidly, apparently faster than a power law; the
interpretation of this behavior is unclear, but probably due to the
nature of the best rational approximants to the incommensurate system
used in the study.]

Finally we compare our present results with those of the random friction
(or ``ratchet'') model \cite{Mihalyetal,Webman87},
which contains randomness and short range
interactions, but {\em no jumps}.

Although this model can be solved exactly only above threshold, much of
the behavior below threshold can be guessed.
A static configuration of the system consists of pinned phases which sit
exactly at the ratchet positions, on which the net force (excluding the
ratchet constraint) is negative, and unpinned regions with zero net force.

The distribution of threshold fields is trivially characterized by an
exponent $ \nu_{\rm T}=2/d$ since $ F_{T}(\{ h_{ i}\})=\left<  h_{ i}
\right>_{ i}$ (where
$\left<\right>_{ i}$ denotes spatial averaging in a given configuration).
Note, however, that the mean $\overline{ F_{T}}(L)$ is size independent.
Right at threshold, the phases are just given in terms of the Fourier
transform $h(q)$ by
 \begin{equation}
\label{eq:rfatft}
\phi(x)=\sum_{q} \frac{-h(q)+\left< h \right>}{q^{2}}
 \end{equation}
so that the rms spread of phases $\sim L^{(4-d)/2}$.
Ignoring the effects of the tail of the distribution (which probably
gives rise to only $\ln(L)$ corrections) this suggests that (for $d<4$)
$P_{T}(L) \sim L^{(4-d)/2}$, so
 \begin{equation}
\label{eq:rfpoldiv}
\frac{\gamma-1}{\nu}=\frac{4-d}{2}.
 \end{equation}
The distribution of linear sizes of unpinned regions will be cutoff above
a correlation length $\xi$ which is the size of the largest regions which
have exceeded their local threshold.
Thus we have
 \begin{equation}
\label{eq:rfnus}
\nu= \nu_{\rm T}=2/d.
 \end{equation}
{}From  Eq.~(\ref{eq:rfpoldiv}) one obtains $\gamma=4/d$ \cite{Webman87},
for a generic approach to threshold.
When the field is decreased from threshold, the ratchet constraints will
immediately repin some regions.
The resulting singularities in $ \chi^{\downarrow}$ will definitely be weaker
than
$ \chi^{\uparrow}$ on a typical approach; concomitantly there will be
characteristic
length $ \nu_{\ell}<\nu$.
Further study of this non-trivial length may be useful even though the
ratchet model, because of the absence of jumps, is very different on
first approach to threshold from the more realistic CDW model studied here.

The values of $\gamma$ we have found in one and two dimensions appear to
be somewhat lower than the ratchet model result $4/d$.
Nevertheless our errors are large enough not to exclude these values.
Well above threshold, the ratchet model has larger deformations than the
CDW model (although they scale with the same exponent
\cite{narayanfisher}).
This may well also be the case at and below thresholds, perhaps leading
to $\gamma<4/d$.

Renormalization group calculations in $d=4-\epsilon$
yield $P_{T}\sim L^{\rho}$ with $\rho=\epsilon/2$ to leading order in
$\epsilon$; this result might hold to all orders in $\epsilon$
\cite{narayanfisher}.

\subsection{Universality and scaling}

A way to generalize the ratchet model to include jumps and make it more
realistic is to replace the cosine potential in the CDW model with a
sawtooth with finite slope, rather than the infinite downward slope
of the random friction model.
A limit of this model, called the ``ratcheted-kick'' model,
has been studied by us above threshold, and it
appears to be in the same universality class (at least as far as the
dominant scaling behavior) as the cosine CDW model in two dimensions and
probably also for $d=1$ and 3.
This is {\em a priori} somewhat surprising since the models yield {\em
different} dynamic exponents both in mean field theory and in zero
dimensions (i.e., finite systems)!

The following conjecture is naturally suggested:
{\em for all properties which involve
the jumps, the cosine and sawtooth models are in the same universality
class in low dimensions}.
Thus, for example, on a generic irreversible approach to threshold,
$\gamma$ and the distribution of avalanche sizes will be the same
\cite{foot:rkcosjumps}.

Quantities which involve smooth evolution rather than jumps must,
however, be different.
Thus $ \gamma_{\ell}$ will be different (or nonexistent) in the sawtooth
model, and there are no obvious analogs of $\mu$ and $\alpha$ in the
sawtooth model (at least providing the modes which go unstable and
trigger avalanches are localized, which, in contrast to the ratchet
model, we expect them to be).
Nevertheless, the density of potentially unstable regions which are
destabilized by a small increase in $\delta f$ might well be the same in
both systems.
In the cosine pinning model,
the density of these regions diverges as $ \left| f \right|^{d \nu_{\ell}-1}
\sim
 \left| f \right|^{(\alpha-1)/2}$ (using $\mu=1/2$) so
that the density of modes
which would be destabilized by decreasing $ \left| f \right|$ to
$ \left| f/2 \right|$ is $ \left| f \right|^{d \nu_{\ell}}= \left| f
\right|^{(\alpha+1)/2}$ (for the cosine
model, this is the number of modes with frequency of order
$ \left| f \right|^{\mu}$ as noted earlier).

We thus conjecture that an exponent equivalent to $d \nu_{\ell}$ or
$(\alpha+1)/2$ can be defined for the sawtooth model and that this will
be the {\em same} as for the cosine model in low dimensions.
This certainly merits direct testing in the sawtooth model.
Although the reversible regime for $F$ decreasing from $ F_{T}$ should also
exist for the sawtooth model, it is not clear whether $ \nu_{\ell}$ can be
found
from finite size corrections to the polarizability, since there are no
obvious precursors of the local instabilities in this model.

We now discuss the conjectured universality in the context of
the distribution of avalanche sizes seen in the cosine model and the
rate at which avalanches are triggered.
An important observation is that the distance between local regions which
would be destabilized by an increase in the local driving force by of
order the distance to the bulk threshold appears to be much less than
the correlation length $\xi$ (since $\xi\sim \left| f \right|^{-\nu}$ and the
distance between destabilized regions behaves as $\sim \left| f \right|^{-
\nu_{\ell}}\ll
\xi$).
This is even more striking if one notes that one should perhaps consider
the effective increase in the force on each region as $F$ is increased to
threshold as being enhanced by the divergent polarization so that one
might consider the number of regions which would go unstable if the local
force were increased by $ \left| f/2 \right|^{1-\gamma}$ rather than just by
$ \left| f/2 \right|$.
This would, however, overestimate the effects since we know (from the
no-passing rule) that at least some region does not move by as much as
$2\pi$ even when $F$ is increased all the way from $ \mbox{$F_{T}^{-}$}$ to $
\mbox{$F_{T}^{+}$}$.

These considerations lead naturally to the consideration of the distribution
of avalanche sizes near threshold.
If the scaling form  Eq.~(\ref{eq-dos-scaling}) for the distribution of
almost
unstable modes is correct in the irreversible approach to threshold, then
the density of avalanches $ n_{\rm av}$
which is triggered by a small increase $\delta f$ is
 \begin{equation}
\label{eq:avadensity}
 n_{\rm av}\delta f \sim  \left| f \right|^{(\alpha-1)/2} \delta f .
 \end{equation}
Our avalanche data for the $256^2$ system of  Fig.~\ref{avalanche} are
consistent with a number of events per unit field diverging roughly as
$ \left| f \right|^{-0.10\pm0.05}$, consistent with
Eq.~(\ref{eq:avadensity}).
If a finite fraction of the avalanches were of size $\sim \xi^{d}$
($\kappa<0$), then
this would result in $\gamma=(1-\alpha)/2+d\nu=2.10\pm0.09$,
which is probably an overestimate.
Indeed, from the discussion in connection with the ratchet model above,
we expect that in $d=2$, $\gamma \le 2 \le d\nu$ (note that for $d=1$,
$\alpha$ may well be negative, as it is for the incommensurate model).
This suggests that the typical avalanche is {\em not} of the size of
the correlation length.

Given a probability
distribution of avalanche sizes $p_{av}(l,f) dl/l$, for avalanches of
linear size $l\sim (\Delta P)^{1/d}$
at reduced field $f$, and the normalization
$\int_{0}^{\infty}p_{av}\,dl/l= n_{\rm av}$, we have
 \begin{equation}
\label{eq:polfromava}
 \left| f \right|^{(\alpha-1)/2}
\int_{0}^{\infty}p_{av}(l,f)l^{d}\frac{dl}{l}
\sim  \left| f \right|^{-\gamma}.
 \end{equation}
This assumes that the typical avalanches are non-fractal,
qualitatively consistent with our numerical results.
A natural scaling form for $p_{av}$, consistent with our data, is
 \begin{equation}
\label{eq:avascale}
p_{av}(l,f)\sim l^{-\kappa}\Phi(l/\xi),
 \end{equation}
with $\Phi(u) \rightarrow{\rm const.}$ for $u\ll 1$ and decaying rapidly for
$u\gg
1$.
In this case, we have
 \begin{equation}
\label{eq:avascalerel}
(d-\kappa)\nu=\gamma-\frac{1-\alpha}{2},
 \end{equation}
suggesting for $d=2$,
 \begin{equation}
\kappa=0.3\pm0.2,
 \end{equation}
consistent with our numerical result of  Sec.~\ref{subsec:avalanches}.
A power law decay of $p_{av}(l,f)$ for $l\ll \xi$ has the natural
interpretation that the probability of an avalanche reaching size $>2l$
{\em given} that it is larger than $l$ is scale invariant.
If it is assumed that the exponent $\kappa$ for
the power law distribution of avalanche sizes at threshold
is independent of the model, as suggested by our agreement with the
distribution in models of self-organized criticality, and that
$\gamma$ is independent of the details of the model, we again reach the
remarkable conclusion suggested above:
that $\alpha$, which gives the divergence
of the avalanche triggering rate as threshold
is approached, is also a universal feature of CDW models.

\subsection{Relation to dynamics above threshold}

In another paper, we will present detailed results on the dynamics of the
cosine and sawtooth models above threshold.
There, the steady state is a {\em unique} periodic function of time with
period $2\pi/v$ \cite{uniqueaam} and
 \begin{equation}
v\sim f^{\zeta}
 \end{equation}
with $\zeta = 0.63\pm 0.06$ for both models in $d=2$ \cite{uniqueaam}.
Again, for the cosine model a double finite size crossover is seen,
analogous to that found below threshold in the linear polarizability.
We note that a corollary of the relations proposed by Coppersmith
and Fisher \cite{SNCandDSF}
in the 1-d incommensurate case is that $\zeta+ \nu_{\ell}=2\mu$.
This relationship clearly does not hold in the $d=2$ random models.
The jumps in the incommensurate model are correlated over the full system
and a picture of locally propagating avalanches is not applicable, so that
the dynamics must be described by a different picture.

For conventional phase transitions, an understanding of the scaling
behavior is greatly enhanced by the addition of an ordering field
which can take the system smoothly from one phase to the other.
Indeed, in cases where such an ordering field {\em per se} does not exist
--- e.g., spin glasses --- the understanding of the transition is far
less complete.

In the case of interest here --- charge density waves and related
problems --- it is not at all clear that a smooth connection between the
two ``phases'' should exist, since they are so radically different, one
involving no motion at all in equilibrium, but with a high degree of
metastability, and the other dynamic, but unique.
One possible way to connect the two phases is by adding thermal noise
which will round out the threshold and yield a finite velocity for any
non-zero force, $F$.
Noise can be added in two rather different ways.
The first is to add a Langevin white noise $\eta( i,t)$ to the
equation of motion, with the variance of $\eta$ proportional to the
temperature $T$.
This will, in the stationary phase, primarily affect the modes which are
near to an instability.
Such an approach has been used before in mean field theory and the
velocity as a function of $T$ and $f$ has been found to exhibit a scaling
form
\cite{dsf:mft}.
Numerical results and arguments based on the distribution of barrier
heights in finite dimensional systems are consistent with this scaling
form \cite{aam:thermal}.
The exponents, however, are dependent on the form of the pinning
potential, as the dependence of barrier heights on
reduced field differs between smooth potentials and those with cusps.
The effects of thermal noise are therefore nonuniversal.

{}From the above discussion of the dominance of jumps over the smooth
motion, it is probably better, for examining universal properties,
to add a noise which will trigger jumps in a way which depends
much less on the details of the potential.
The hope is that the sawtooth and cosine models will then behave
similarly.
We thus consider giving random ``kicks'' to individual phases with fixed
impulse magnitude (of order $\pi$) at a slow rate $\Theta$.
For any finite $\Theta$, the mean velocity will be non-zero.
Below threshold in the limit of infinitesimal $\Theta$, the mean velocity
$\left< v \right>$ should be proportional to $\Theta$ (provided the kicks
are large enough) so that we may define a linear response
 \begin{equation}
\Xi\equiv \left.\frac{d\left< v \right>}{d\Theta}
\right|_{\Theta\downarrow 0}.
 \end{equation}
Near to threshold, $\Xi$ will presumably diverge, and above threshold,
$\left< v \right>$ will be non-zero even at $\Theta=0$, but we would
again expect $\Xi$ to be finite.
Thus $\Xi$ is somewhat like an order parameter susceptibility near a
conventional thermal transition.
Although we must leave investigation of this kind of noise response for
future investigation, a few  remarks relevant to the present paper are in
order.

The main effects which cause subtleties below threshold are transients
and the non-uniqueness in the absence of noise.
A natural way to define, at least statistically, another type of
preferred configuration at a given $F< F_{T}$ is to turn on a very small
noise, let the system equilibrate (if the kicks are large enough, the
steady state distribution will presumably be unique, although this needs
establishing), turn off the noise, let the system relax, and then study
the statistical behavior of the resulting static configurations (e.g.,
polarizability, distribution of modes, or closeness to local
instabilities, etc.).
It is by no means clear, {\em a priori}, that such a procedure --- which
should be qualitatively similar to that in real experiments --- will
produce configurations which are similar to {\em either} of the histories
we have studied in this paper.
If not, then it is perhaps only {\em transients} in the behavior above
threshold and not the steady state itself which could have critical
behavior related to that of, say, the irreversible approach to threshold.
If, on the other hand, the configurations are statistically similar to
those produced on generic noiseless approaches to threshold, then there
will presumably be dynamic responses in the moving phase --- such as
$\Xi$ --- which can be related to exponents below threshold.

We leave these and related intriguing questions for future study.

The recent renormalization calculations of the dynamics above
threshold suggest that the role of the various correlation lengths is
rather different: above threshold the dominant correlation length
exponent for dynamic quantities in the steady state is
$\nu_+=1/2$ exactly.
A larger exponent $ \nu_{\rm T}=2/d+O(\epsilon^2)$ (and perhaps with no
perturbative corrections) appears and controls the distribution of
threshold fields.
Thus the dominant length below threshold, for irreversible approaches,
diverges with an exponent $\nu_- =  \nu_{\rm T}$, while above threshold
dynamics is controlled by $\nu_+< \nu_{\rm T}$.
Whether this difference is primarily due to the increased level of
cooperativity in steady state above threshold, or to some other reason,
is unclear.
In addition, whether or not an exponent equal to $ \nu_{\rm T}$ might control
the dynamics far from steady state above threshold is also interesting.

Another set of questions concerns relationships to other so-called
``self-organized critical'' transport \cite{marg-stable},
which we have discussed
above only as far as it relates to the behavior at threshold and as
threshold is approached.
It is plausible that such relations can also be developed in the
sliding state.
If CDW's are driven at constant, very slow, {\em current} (i.e., fixed mean
velocity) by an external field
--- somewhat analogous to the quasistatic limit of ``sandpile''
dynamics that has been extensively studied recently --- then the system
is near criticality and will exhibit power law correlations, etc.
How the other problems studied are related to CDW's (except in spirit
and the quantitative similarities in the distribution of avalanches)
is an open question: ``sandpile'' models typically have thermal noise
(analogous to small $\Theta$ above) but no quenched randomness.
Models of earthquakes \cite{Carlson-Langer}
with inertia but no quenched randomness have been
studied, and so have models with intrinsic randomness, but no inertia.
Which are more realistic is controversial.

At this point, perhaps all that can safely be said is that the
relationships between different non-linear collective transport phenomena
and the possible existence of some degree of universality are likely to
remain challenging problems for some time.

\underline{Acknowledgments:} We thank P.~B.\ Littlewood,
S.~N.\ Coppersmith, P.\ Sibani, T.\ Hwa, C.\ Myers, and O.\ Narayan for
useful discussions.
DSF is supported by the A.~P.\ Sloan Foundation and the National
Science Foundation through Grant DMR-9106237 and via the Harvard
University Materials Research Laboratory.
This research was conducted using the computational resources of
Argonne National Laboratory and the Northeast Parallel Architectures
Center (NPAC) at Syracuse University.

\newpage
\begin{table}
\squeezetable
\caption{Numerical results for critical exponents of the charge-density
wave model defined by  Eq.~(\ref{eq:motion}).}
\begin{tabular}{clr@{ }c@{}lr@{ }c@{}l}

Exponent&Definition&\multicolumn{3}{c}{$d=1$}&\multicolumn{3}{c}{$d=2$} \\
\tableline
$\gamma$ & Divergence of total polarizability (irreversible
path),  Eq.~(\ref{gammadef})
& \dec 3.0 & $\pm$ & \dec 0.5 & \dec 1.8 & $\pm$ & \dec 0.15 \\

$\rho$   & Size dependence of configuration width at threshold,
	 Eq.~(\ref{rhodef})
& \dec 1.3 & $\pm$ & \dec 0.3 & \dec 0.8 & $\pm$ & \dec 0.2 \\

$ \nu_{\rm T}$   & Size dependence of threshold field distribution,
	 Eq.~(\ref{nutdef})
& \dec 2.01 & $\pm$ & \dec 0.02 & \dec 1.01 & $\pm$ & \dec 0.03 \\

$ \nu_{\rm n}$   & Finite-size crossover field of polarization,
Eq.~(\ref{eq:pscale})
& \dec 2.0 & $\pm$ & \dec 0.5 & \dec 1.0 & $\pm$ & \dec 0.1 \\

$\kappa$  & Avalanche size distribution near threshold,  Eq.~(\ref{kappadef})
& & --- & & \dec 0.34 & $\pm$ & \dec 0.10 \\

$ \gamma_{\ell}^{R}$& Cusp in linear polarizability (reversible path),
	 Eq.~(\ref{gammalrdef})
& & --- & & \dec -0.42 & $\pm$ & \dec 0.05 \\

$ \gamma_{\ell}^{I}$& Cusp in linear polarizability (irreversible path),
	 Eq.~(\ref{gammalidef})
& & --- & & \dec -0.40 & $\pm$ & \dec 0.12 \\

$\alpha$  & Distribution of linear eigenvalues at threshold,
	 Eq.~(\ref{eq-dos-scaling})
& & --- & & \dec 0.84 & $\pm$ & \dec 0.12$^{\rm a}$ \\

$ \nu_{\ell}$    & Finite size crossover of linear polarizability
(reversible path),  Eq.~(\ref{nuldef})
& & --- & & \dec 0.44 & $\pm$ & \dec 0.08 \\

\end{tabular}
\label{table1}
\tablenotes{$^{\rm a}$ Calculated using  Eq.~(\ref{eq-gamma-scaling})}
\tablenotes{}
\end{table}


\figure{\label{nopassing}
A schematic illustration in one dimension of the ``no-passing'' rule for CDW
configurations.
The lines show the phases, $ \varphi_{ i}$, as a function of position, $ i$,
for two configurations, one of which (open circles) initially trails
(is less than) the other (solid circles) at time $t=0$.
Both configurations are driven by the same external field $F(t)$.
As the configurations evolve from their initial positions,
according to the equations of motion  Eq.~(\ref{eq:motion}), they may come
close to intersecting, as shown in the figure for $t>0$.
They never cross, though: as the two configurations approach
each other at some site, the drive and pinning forces on the phase at that
site tend to cancel, but the elastic forces, which tend to flatten out the
configuration, do not allow the configurations to pass through each other.
The arrows indicate the relative elastic forces for $t>0$.}

\figure{\label{leastconfig}
An illustration of the partial ordering of the configurations in the
static state.
The lines show the phases $ \varphi_{ i}$ as a function of position in
the lattice, $i$, for static solutions to the equations of motion for
the CDW.
The lowest line shows the initial configuration $\{ \varphi_{ i}(0)\}$ static
at field
$F(0)$, while the other lines show configurations
$\{ \varphi_{ i}^{*}\}\in{\cal A}^{*}( \varphi_{ i}(0))$ which is the set of
configurations with $ \varphi_{ i}^{*} \geq  \varphi_{ i}(0)$,
for all $ i$, that are static at the field $F^{*}$,
where $F\leq F^{*}\leq \mbox{$F_{T}^{+}$}$.
Given the initial configuration $\{ \varphi_{ i}(0)\}$, if
the field is raised to $F^{*}$, the configuration must
converge to the unique configuration, $\{ \varphi_{ i}^{\infty}\}$, shown as
the heavy
line, which is the {\em lowest} stationary
configuration that is {\em above} the initial
configuration.
The configurations that are static at $F^{*}$ and exceed the initial
configuration may cross each other as shown, but no configuration in
${\cal A}^{*}$ may cross the configuration $\{ \varphi_{ i}^{\infty}\}$.}

\figure{\label{hystloop}
Plot of the polarization, $P$, vs.\ applied field, $F$, for a
two-dimensional system of size $64^{2}$.
The initial configuration is the configuration static at the negative
threshold field, $ \mbox{$F_{T}^{-}$}$.
The applied field is increased to the upper threshold value, $
\mbox{$F_{T}^{+}$}$,
and then lowered again.
This particular history of applied field, for an adiabatic variation of
the field, defines two approaches to threshold: the initial path in
configuration space from $ \mbox{$F_{T}^{-}$}$ to $ \mbox{$F_{T}^{+}$}$ is
the ``irreversible''
path and the path for decreasing field, near
$ \mbox{$F_{T}^{+}$}$, defines the ``reversible'' path.}

\figure{\label{threshPvsize}
The polarization at threshold, $P_{T} = P( \mbox{$F_{T}^{+}$}) - P(
\mbox{$F_{T}^{-}$})$,
plotted as a function of the size of the system $L$ for one and two
dimensions.
Straight lines showing power law behavior $P_{T} \sim L^{\rho}$ with
$\rho= 1.3\pm 0.3$ and $ 0.8\pm 0.2$ are shown,
corresponding to the estimated exponents
for one and two dimensions, respectively.}

\figure{\label{pol_one}
Polarization $P$ vs.\ reduced field $f$ for one-dimensional lattice CDW
systems, for the irreversible approach to threshold.  The sample sizes (and
number of realizations averaged over) are indicated.
The dashed line shows power law
behavior  $P \sim  \left| f \right|^{-\gamma+1}$ with
$\gamma= 3.0\pm 0.5$.}

\figure{\label{pol_two}
Polarization $P$ vs.\ reduced field $f$ for the irreversible approach to
threshold in two dimensions.  The system sizes (number of realizations) are
indicated.
The straight line shows power law
behavior, $P \sim  \left| f \right|^{-\gamma+1}$, with
$\gamma= 1.8\pm 0.15$ determined by the fits to the
polarizability shown in  Fig.~\ref{chi-two}.}

\figure{\label{chi-two}
Polarizability $ \chi^{\uparrow}(f)$ for the irreversible path in two
dimensions, which is found by
calculating the numerical derivative of the data of  Fig.~\ref{pol_two},
i.e.,
the difference of the polarization between two consecutive data points.
The straight line shows a fit to the form $ \chi^{\uparrow} \sim  \left| f
\right|^{-\gamma}$,
with $\gamma= 1.8\pm 0.15$.}

\figure{\label{polthreed}
The polarization $P$ as a function of the magnitude of
the reduced field $ \left| f \right|$ on a irreversible approach to threshold
(see text),
for single samples in $d=3$ of size $64^3$ and $128^{3}$.
The critical behavior is difficult to
determine with confidence, but note that the polarization
exceeds $2\pi$ only for fields within $0.1\%$ of threshold.
The dotted line shows an exponent $\gamma=0.33$ for comparison only.
}

\figure{\label{micro}
A fine scale plot of the polarization $P$ (upper curve, right scale)
and linear polarizability $ \chi_{0}$ (lower curve, left scale) for
the irreversible path (increasing $F$)
for a single system of size $128^{2}$.
The discontinuities in the polarization are due to the jumps that occur when
a local minimum of the potential vanishes.
The corresponding spikes in $ \chi_{0}$ are due to the diverging linear
response as these jumps are approached (the divergence has been cut off at
an arbitrary value for the plot).
{}From the size of the jumps in the polarization, it can be seen that, for
reduced fields of this order, the jumps involve only a few of the
degrees of freedom.
In an infinite system, the jumps occur at a dense set of fields, but the
fraction of the field range in which the linear polarizability is
affected by the divergences goes to zero as the system size approaches
infinity.
The dotted line indicates an envelope function, which, in the
infinite volume limit, gives $ \chi_{0}(F)$ with probability one.
}

\figure{\label{dchidf}
Plot of the derivative of the linear polarizability,
$d \chi_{0}(f)/dF$, vs.\ reduced field $f$ for two-dimensional systems
of various sizes, for the reversible approach to threshold.
The fit indicated by the slope of the dashed line
gives $d \chi_{0}(f)/dF \sim f^{- \gamma_{\ell}^{R}-1}$ with
$ \gamma_{\ell}^{R}= -0.42\pm 0.05$.}

\figure{\label{chiT-chi}
Plot of $ \chi_{T}- \chi_{0}(f)$ for reversible (open symbols) and
irreversible
(closed symbols) approaches to threshold in two dimensions.
The threshold polarizability
$ \chi_{T}$ is calculated from the data for $d \chi_{0}/dF$ along
the reversible path.  Straight lines
show fits to cusp-like behavior
$ \chi_{0}(f) \simeq  \chi_{T} - A^{(R,I)}f^{- \gamma_{\ell}^{(R,I)}}$, where
the reversible
path exponent, $ \gamma_{\ell}^{R}$, is found to be $ -0.42\pm 0.05$
and the irreversible path exponent
$ \gamma_{\ell}^{I}= -0.40\pm 0.12$.
Error bars indicate
statistical uncertainties for the reversible path and the uncertainty in
$ \chi_{T}$ for data on the irreversible path;  statistical errors for the
irreversible path are of the order of the fluctuations about the fit.}

\figure{\label{lowmodes}
Plot of the square of the linear eigenvalues, $ \Lambda_{m}^{2}$, vs.\
reduced
field $f$ for the lowest eigenmodes $m=0,1,2$,
in $d=2$ on the reversible approach to threshold.
The straight lines show fits $ \Lambda_{m}^{2}\sim(f_{m}^{c}-f)$.
The extrapolations shown for $f>0$, which determine $f_{m}^{c}$, are not
physical, as once the lowest mode becomes unstable ($\Lambda_{0}=0$),
the CDW is in the sliding state.
}

\figure{\label{dospict}
Scaling picture of the density of states $\rho(\Lambda)$.
The dashed line shows the density of states for the threshold configuration,
which behaves as a power law $\rho \sim \left| \Lambda \right|^{\alpha}$
for small $\Lambda$.
The upper curve,
diverging at finite $\Lambda$ and vanishing for smaller $\Lambda$,
gives the density of states for the reversible path,
while the lower curve, linear at the origin, shows the density of states
for the irreversible path.
Though the distributions for the two histories differ in shape, they share a
common scaling form,
$\rho(\Lambda) \sim \Lambda^{\alpha}  \hat{\rho}(\Lambda/f^{\mu})$, with
the characteristic scale $f^{\mu}$ (indicated by the
vertical dotted line) the same for both histories.}

\figure{\label{Ft}
Average threshold field, $ F_{T}$, as a function of linear size, $L$,
in one and two dimensions.  The error bars show the statistical error in the
average of the threshold field;  the width of the distribution of threshold
fields at each size is larger (see  Fig.~\ref{DFt}.)}

\figure{\label{typFtdist}
Distribution of threshold fields calculated for $128$ samples of
linear size $L=32$ in two dimensions (solid line).
For comparison, a Gaussian fit is shown (dotted line).}

\figure{\label{DFt}
A plot of the width of the threshold-field distribution, $ \mbox{$\Delta
F_{T}(L)$}$,
vs.\ linear size, $L$, in one and two dimensions.
Lines show least square fits to the data for $L\ge 16$.
{}From the slopes of these lines, we find values for the
finite-size-scaling exponent $ \nu_{\rm T}$ of $ 2.01\pm 0.02$ and
$ 1.01\pm 0.03$ in one and two dimensions, respectively.}

\figure{\label{polscl2}
Scaling plot of $P \left| f \right|^{\gamma-1}$vs.\ $L \left| f \right|^{
\nu_{\rm n}}$ for the
polarization of two-dimensional systems, for best fit values of
$ \nu_{\rm n}= 1.0\pm 0.1$ and $\gamma=1.8\pm0.1$.
The deviations apparent at
large $L \left| f \right|^{ \nu_{\rm n}}$ occur for large reduced fields
$1.0> \left| f \right|>0.2$,
where corrections to scaling are pronounced.}

\figure{\label{polscl1}
Plot of $P \left| f \right|^{\gamma}$ vs.\ $L \left| f \right|^{ \nu_{\rm
n}}$ for the
one-dimensional lattice CDW model, with exponents $ \nu_{\rm n}= 2.0$ and
$\gamma=3.5$.  No choice for these exponents gives a single curve
which fits all of the data.}

\figure{\label{avalanche}
Individual avalanche events, indicated by avalanche size $ \Delta P$ on
a logarithmic scale and the field $F$ at which the avalanche occurs,
for a system of size $256^{2}$ on the path $\{  \varphi_{
i}^{\uparrow}{F}\}$; each point
corresponds to a single event.
The solid line indicates a size dependence $\sim( \mbox{$F_{T}^{+}$}-F)^{-d
\nu}$,
with $\nu=1.0$.
The region outlined by the dashed line indicates the events used to
determine the near-critical avalanche size distribution in
Fig.~\ref{avdist}.
}

\figure{\label{avdist}
Logarithmically binned distribution $N(\Delta P)$
of avalanche sizes $\Delta P$ for the events in the near-critical region
indicated by the dashed line in  Fig.~\ref{avalanche}.
The solid line shows the fit $N( \Delta P)\sim  \Delta P^{-\kappa/d}$, for
$\kappa= 0.34\pm 0.10$.
}

\figure{\label{dchidfscl}
Scaled plot of $d \chi_{0}(f)/dF$ for the reversible path in two-dimensional
systems, using the best-fit finite-size scaling exponent $ \nu_{\ell}=0.44$.
Representative error bars show the statistical uncertainty in the scaled
$d \chi_{0}(f)/dF$; for $ \left| f \right|L^{1/ \nu_{\ell}} > 100$, the error
bars are smaller
than the symbol size.
}


\begin{references}
\bibitem{twocoords}
In two dimensions, it is possible to define a potential with both
``troughs'', which a particle can be driven along indefinitely by a field,
and ``pits'', where the particle will be trapped, at the same drive field.
Such a potential can be used for many degrees of freedom to also obtain
a hysteretic velocity-force relationship.
\bibitem{footc}
For the value of $h=5$ for $d=2$, the constant used is $c=3$ and the
additional condition is imposed that no phase may increase by more
than $0.3$ in a single iteration.
\bibitem{foot1}
This equation is {\em not} a diffusion equation in a random potential,
since the $ \varphi_{ i}$ are strongly correlated.  Since there are negative
values for $h\sin( \varphi_{ i}- \beta_{ i})$, a uncorrelated distribution of
these
coefficients in the diffusion equation would lead to ``band tails'',
with positive eigenvalues, implying the instability of the solution.
\bibitem{caveat}
The divergence of the polarizability due to finite-size effects,
discussed below, implies that the limit $L \rightarrow\infty$ must be taken
before the $F\nearrow \mbox{$F_{T}^{+}$}$ limit.
It is assumed here that this ordering of the limits does not
affect the convergence to unique distribution of linear eigenmodes
at threshold, also discussed below.
\bibitem{foot:smallh}
For small $h_{0}$, there is no reversible regime in mean field theory;
however, there appears to be a pathology at this limit.
\end{references}
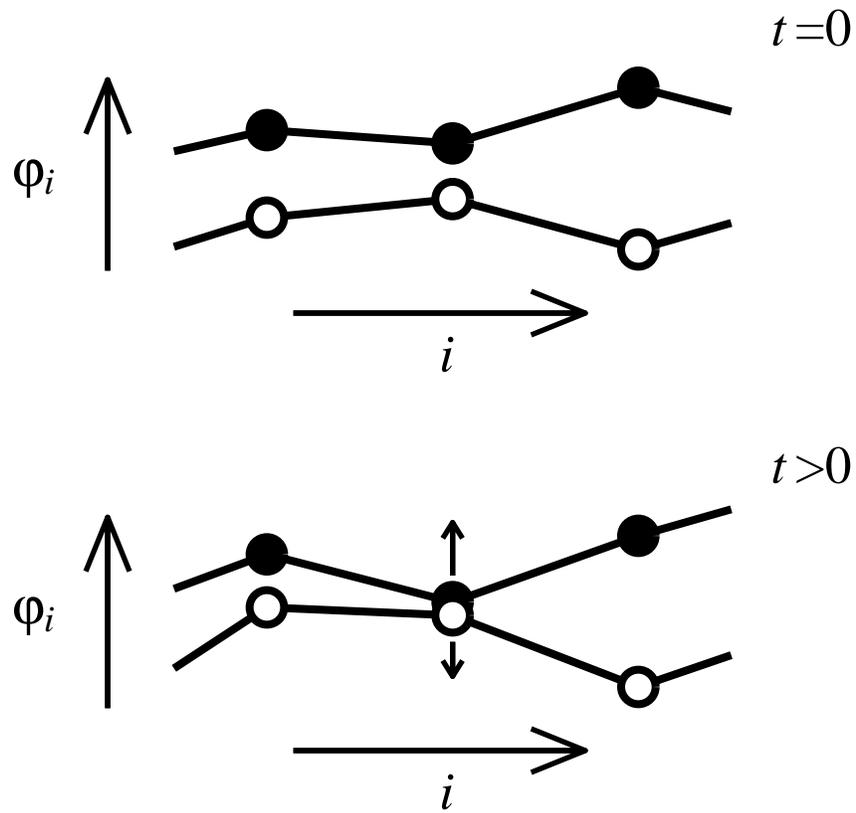
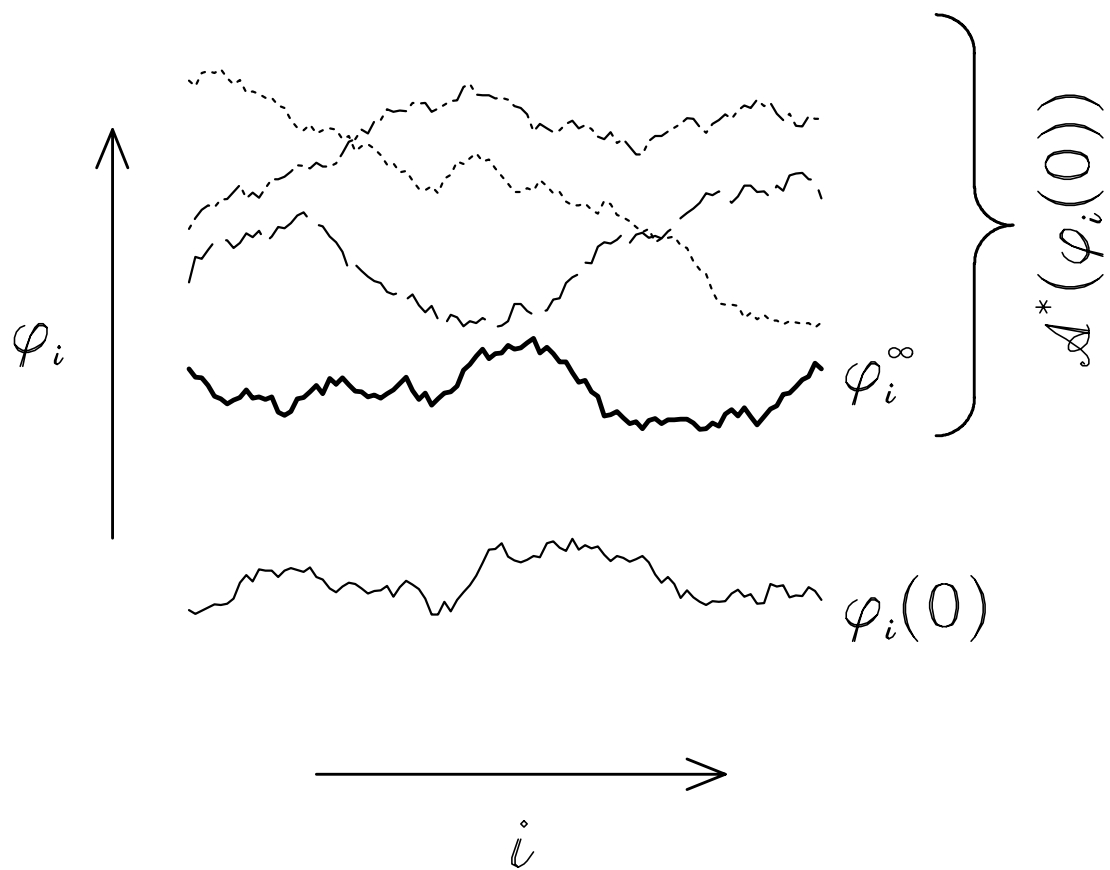
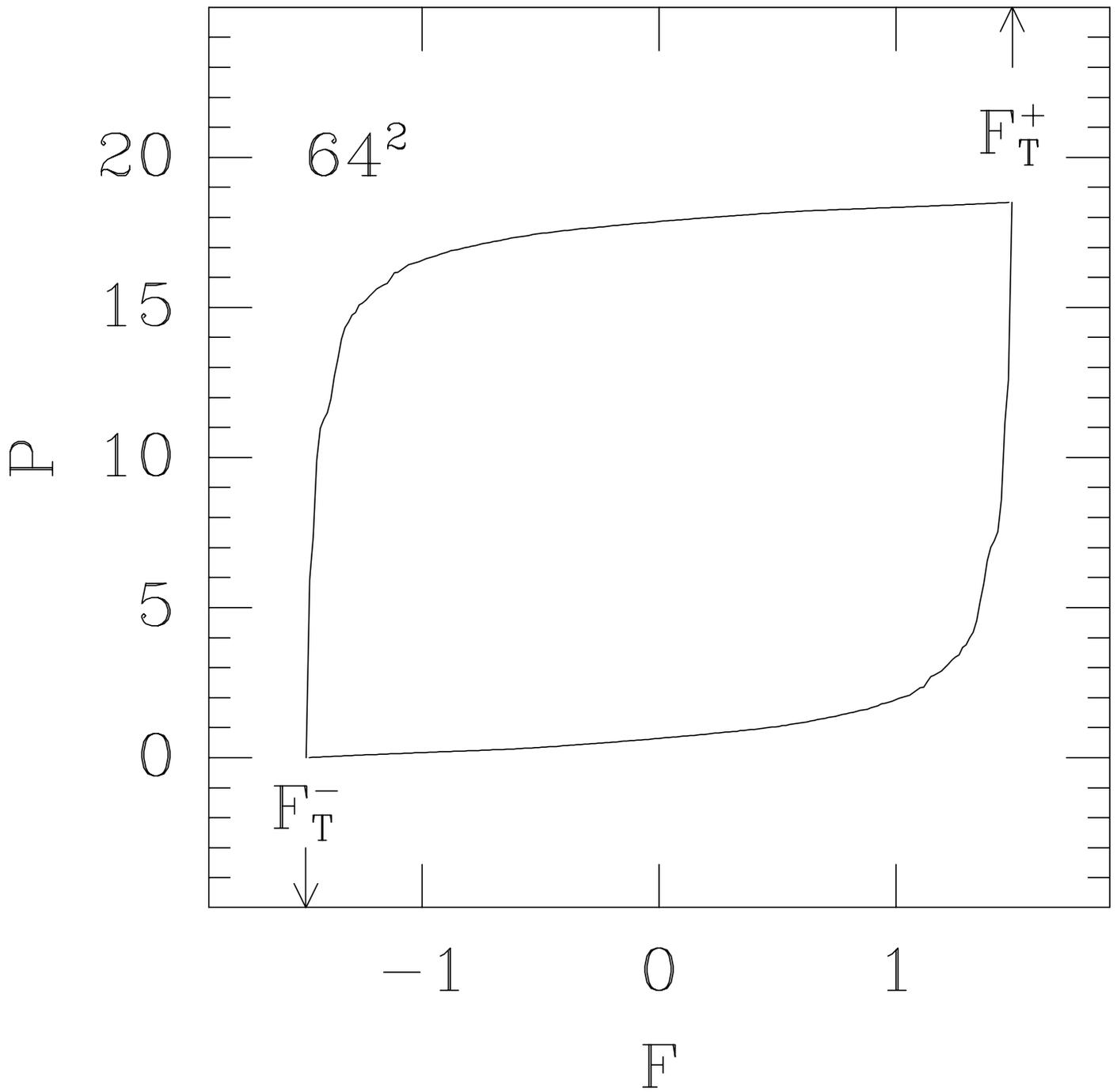
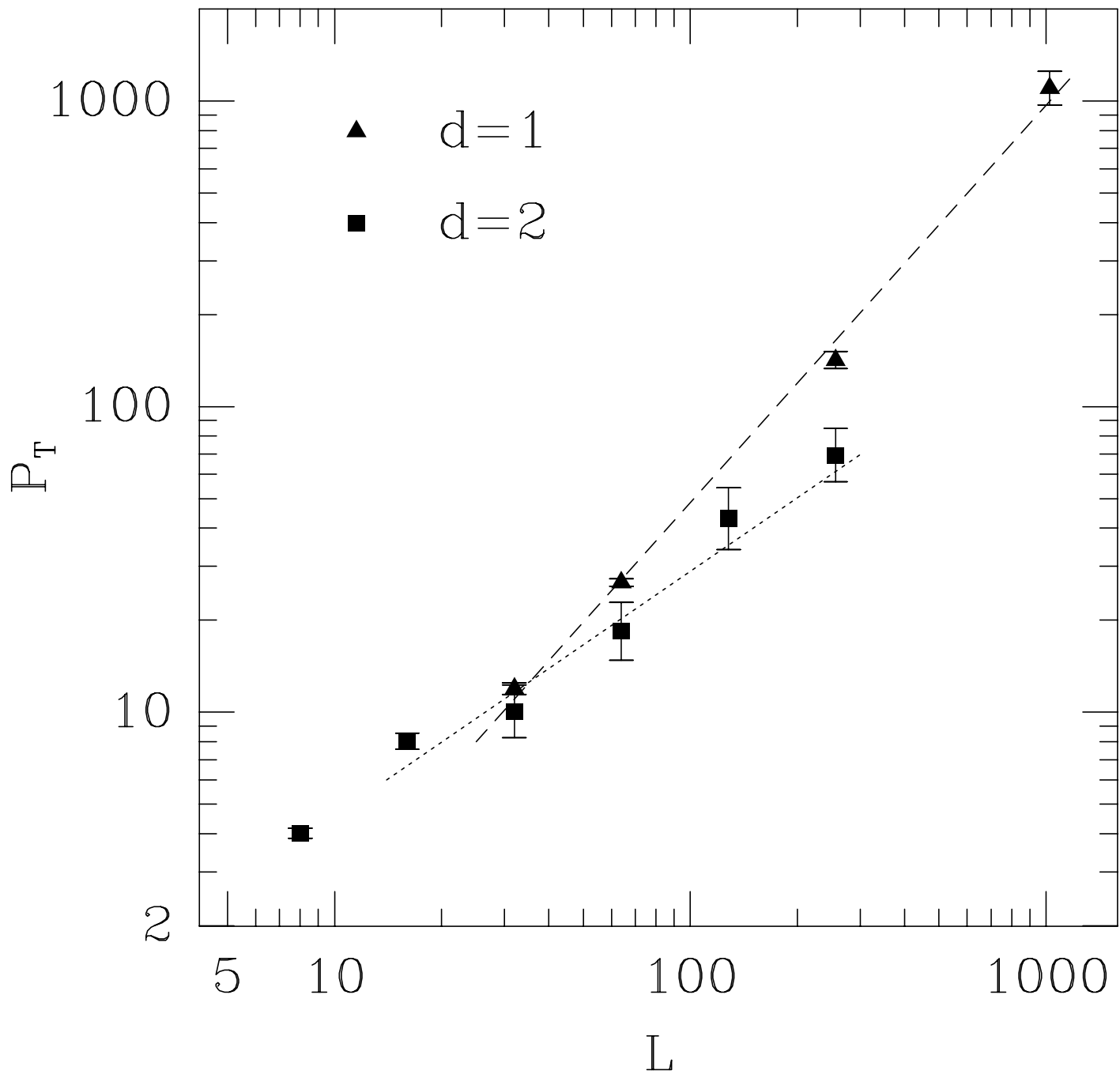
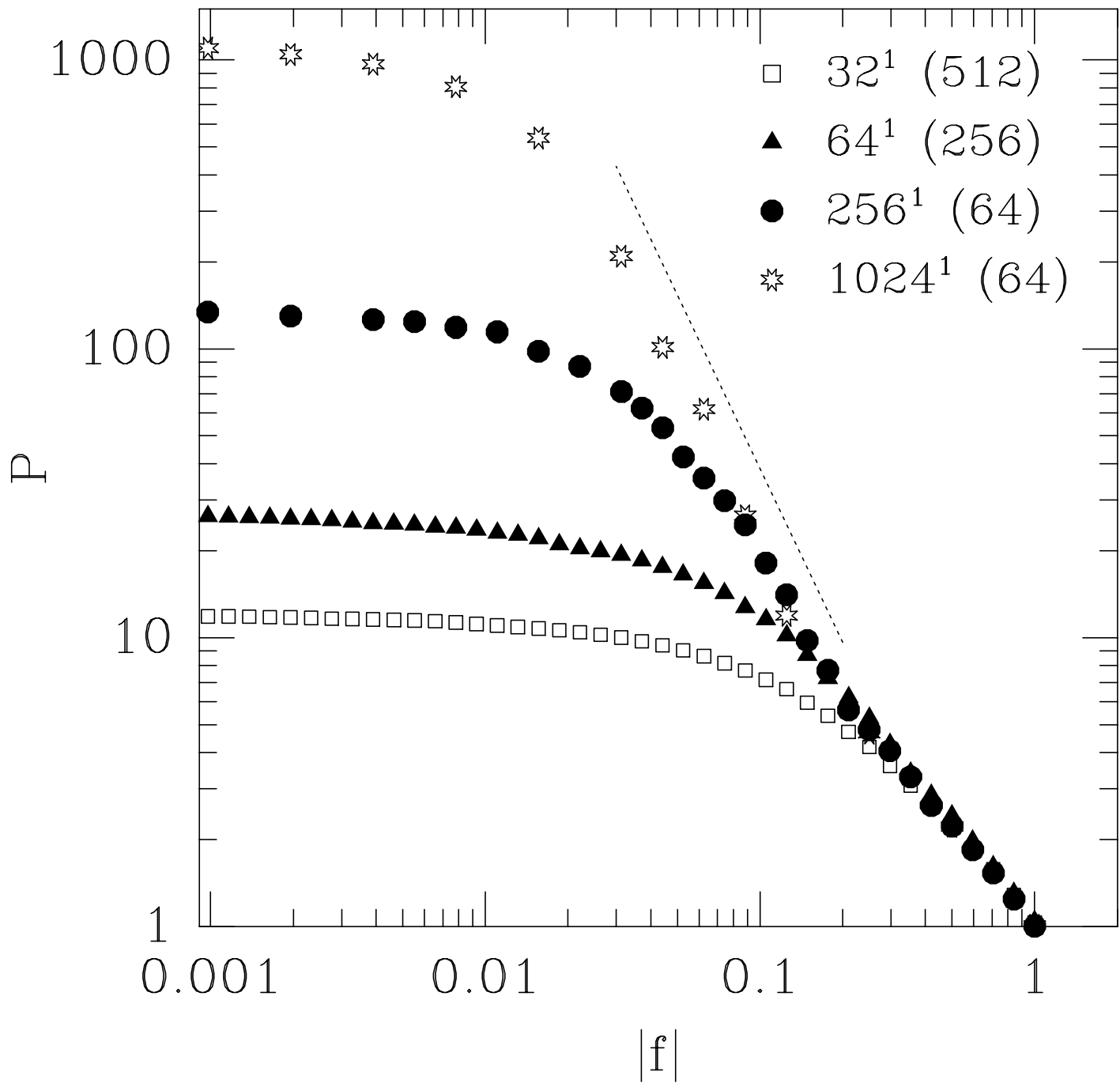
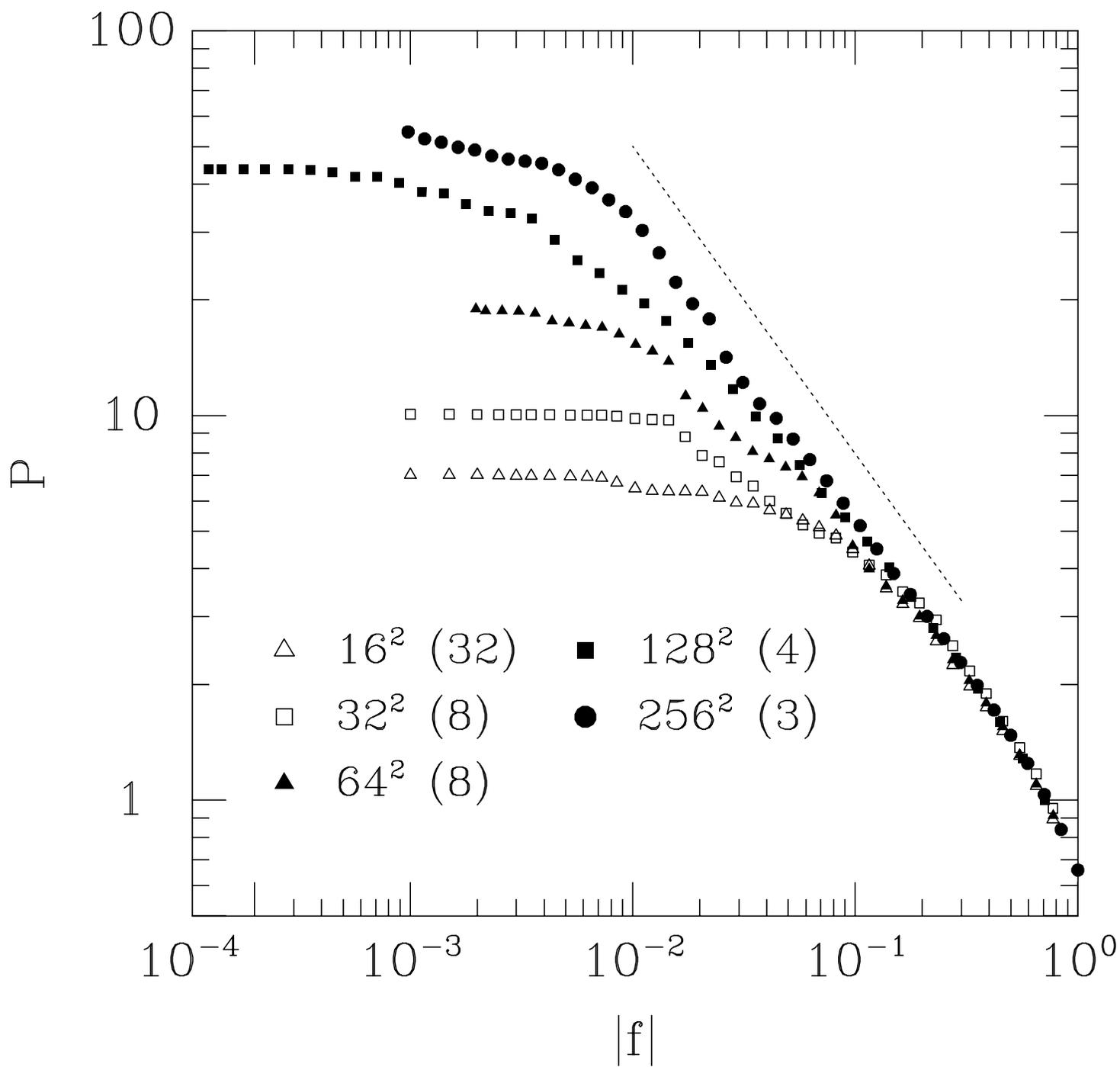
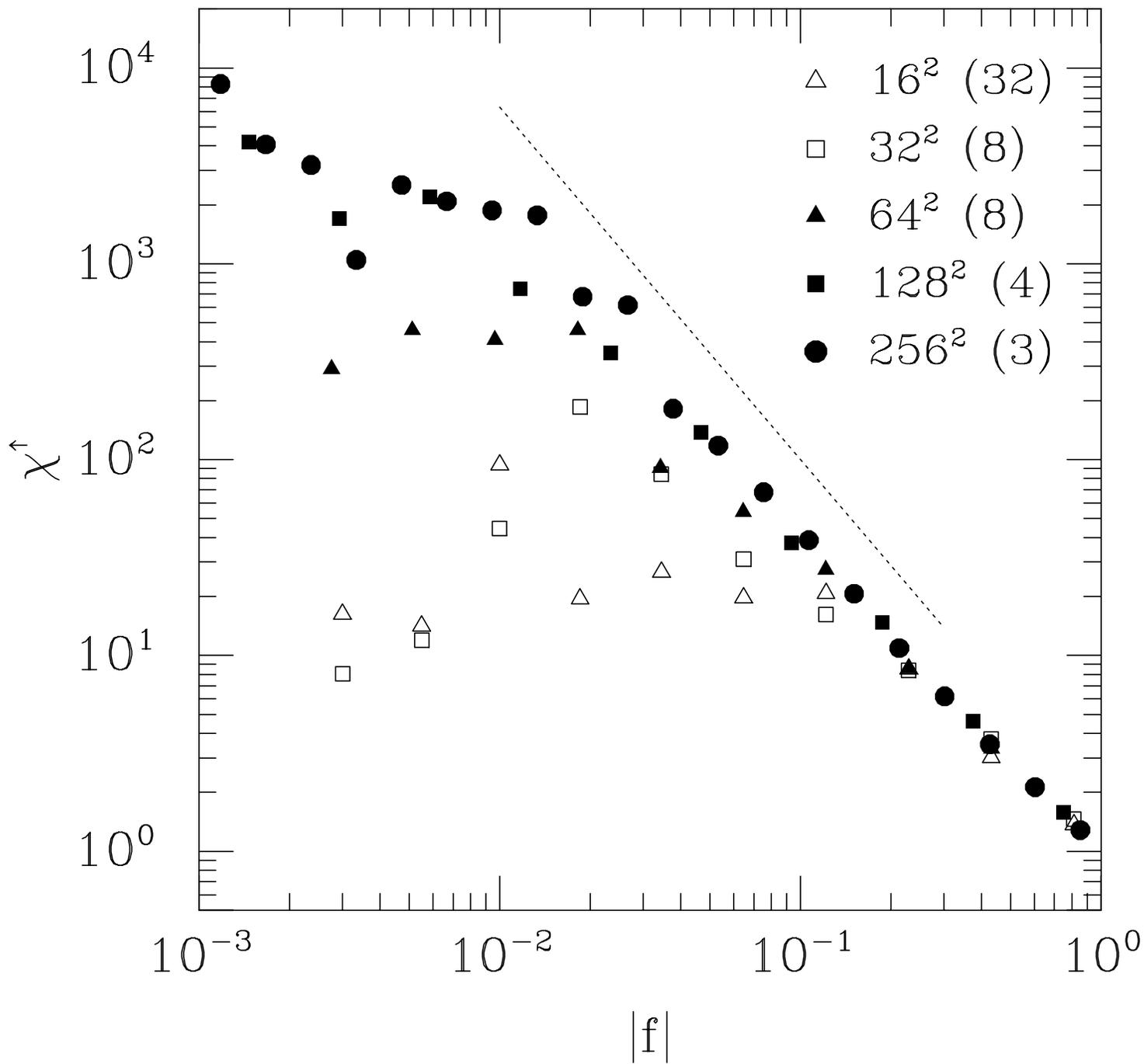
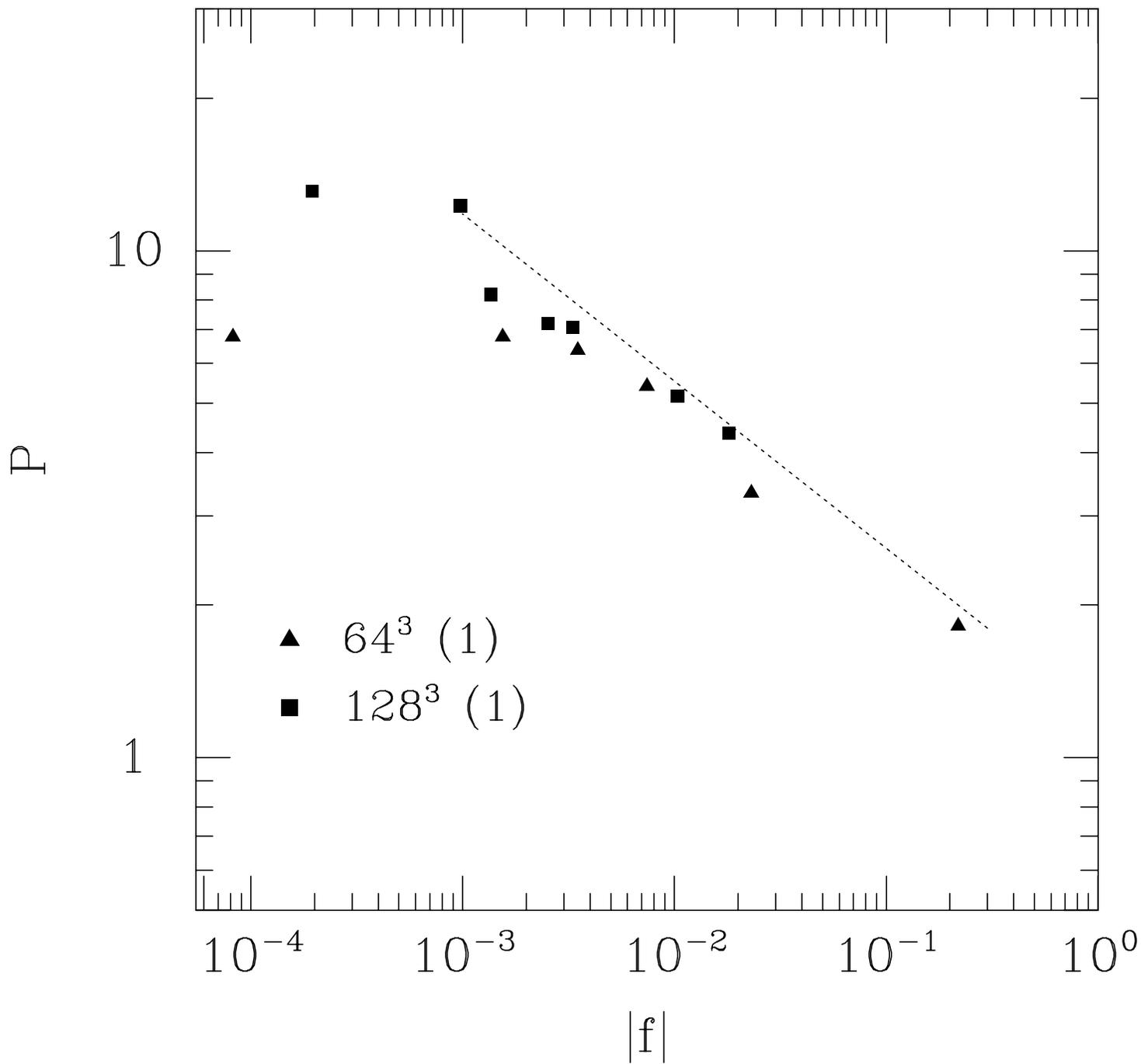
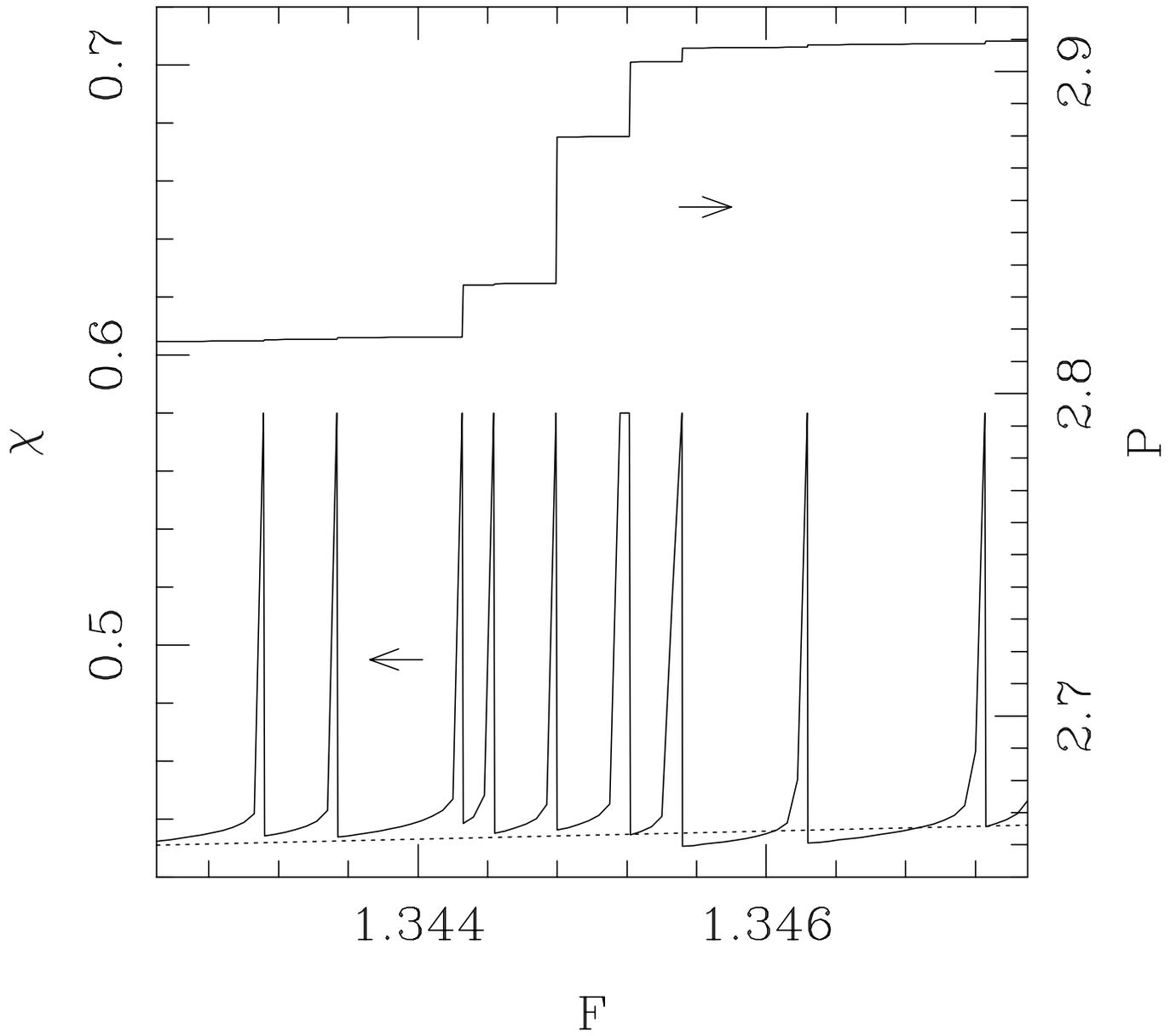
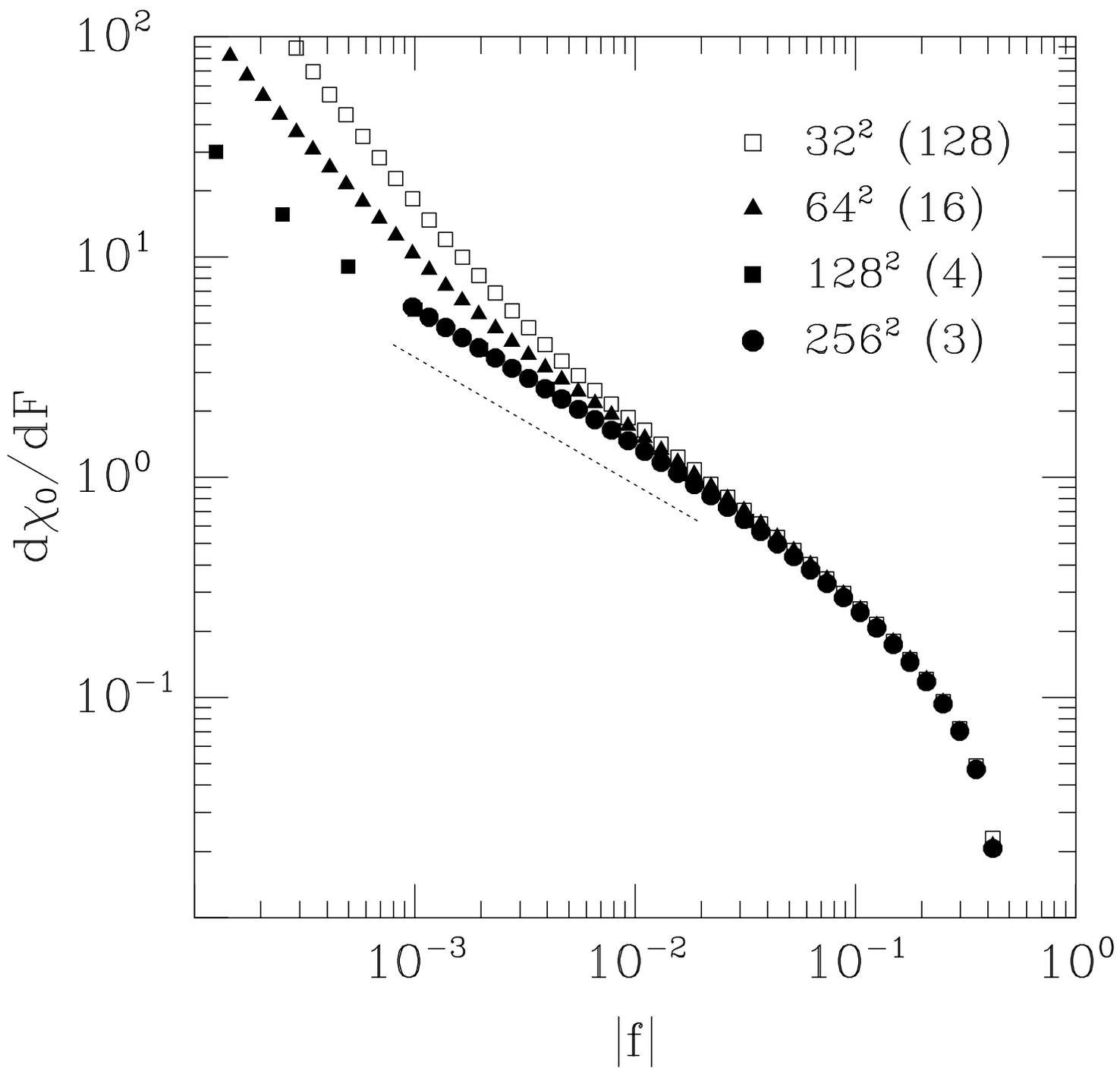
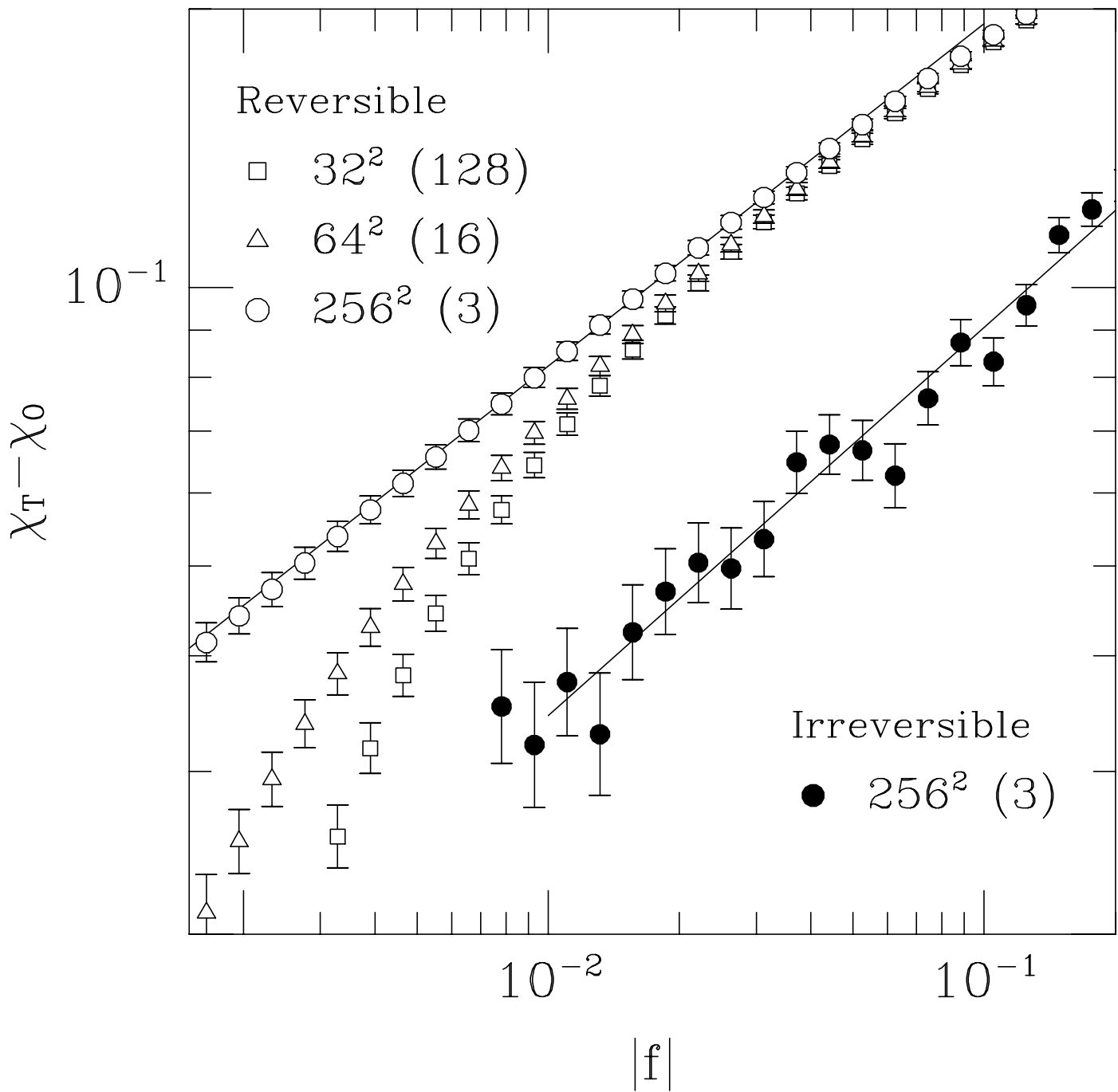
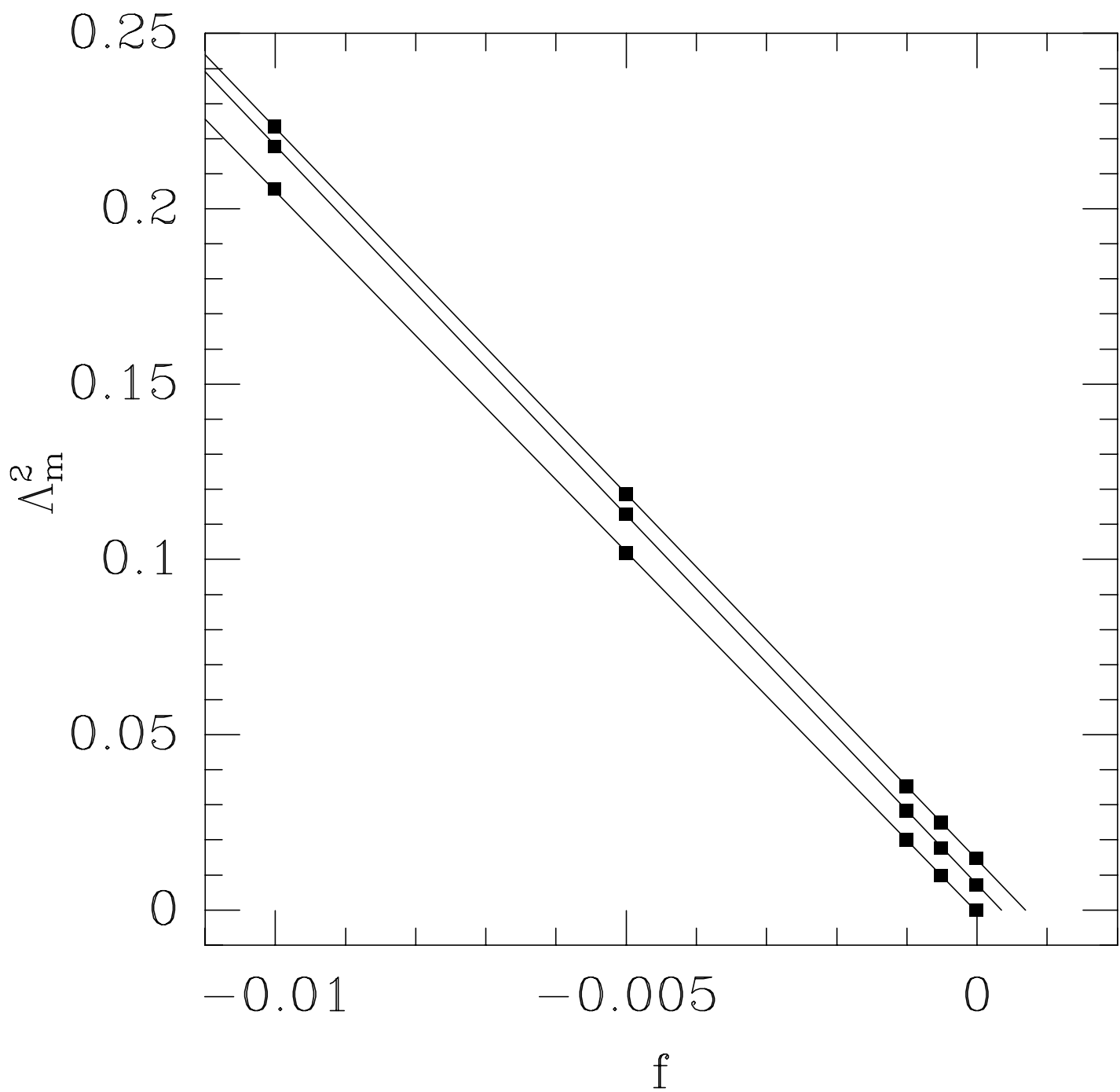
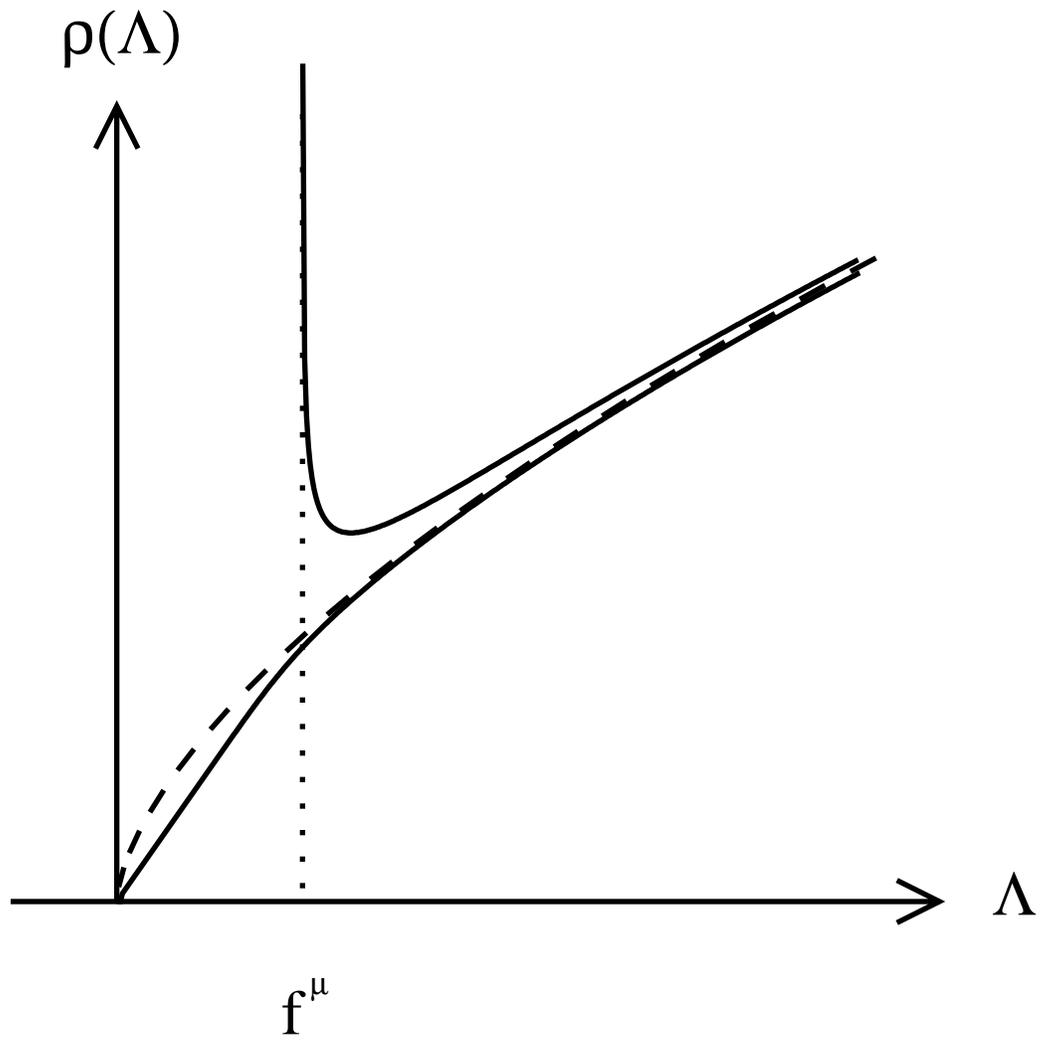
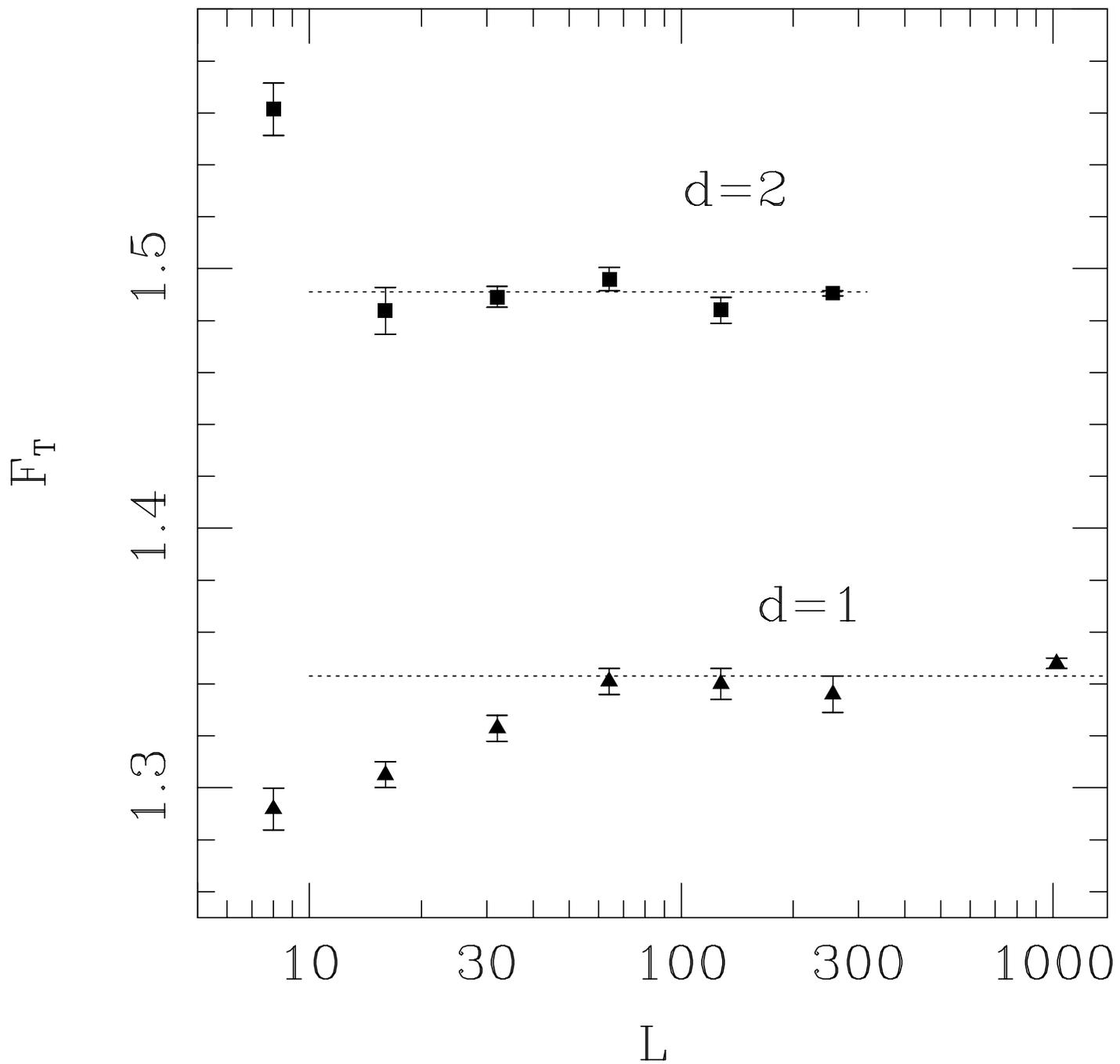
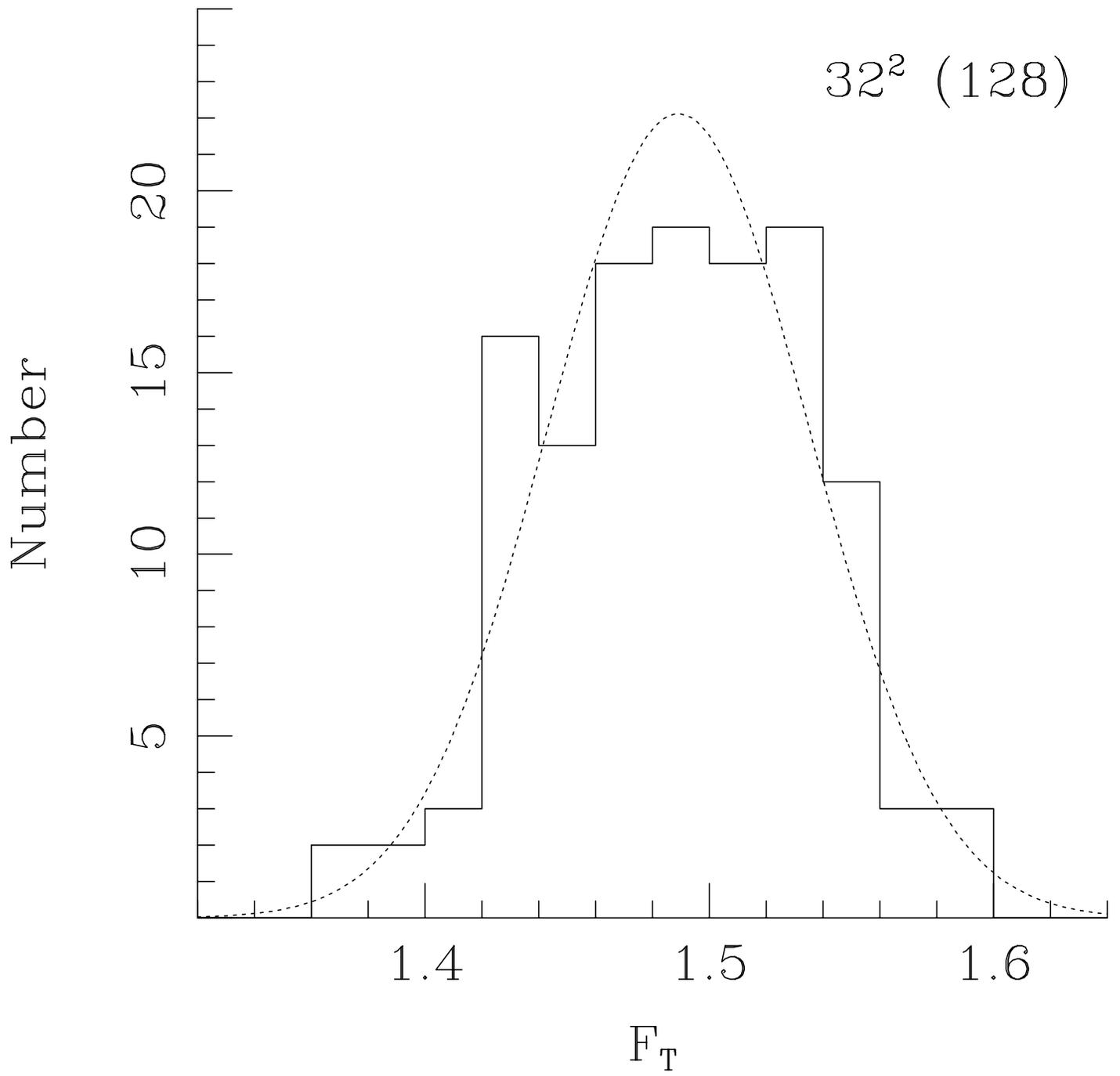
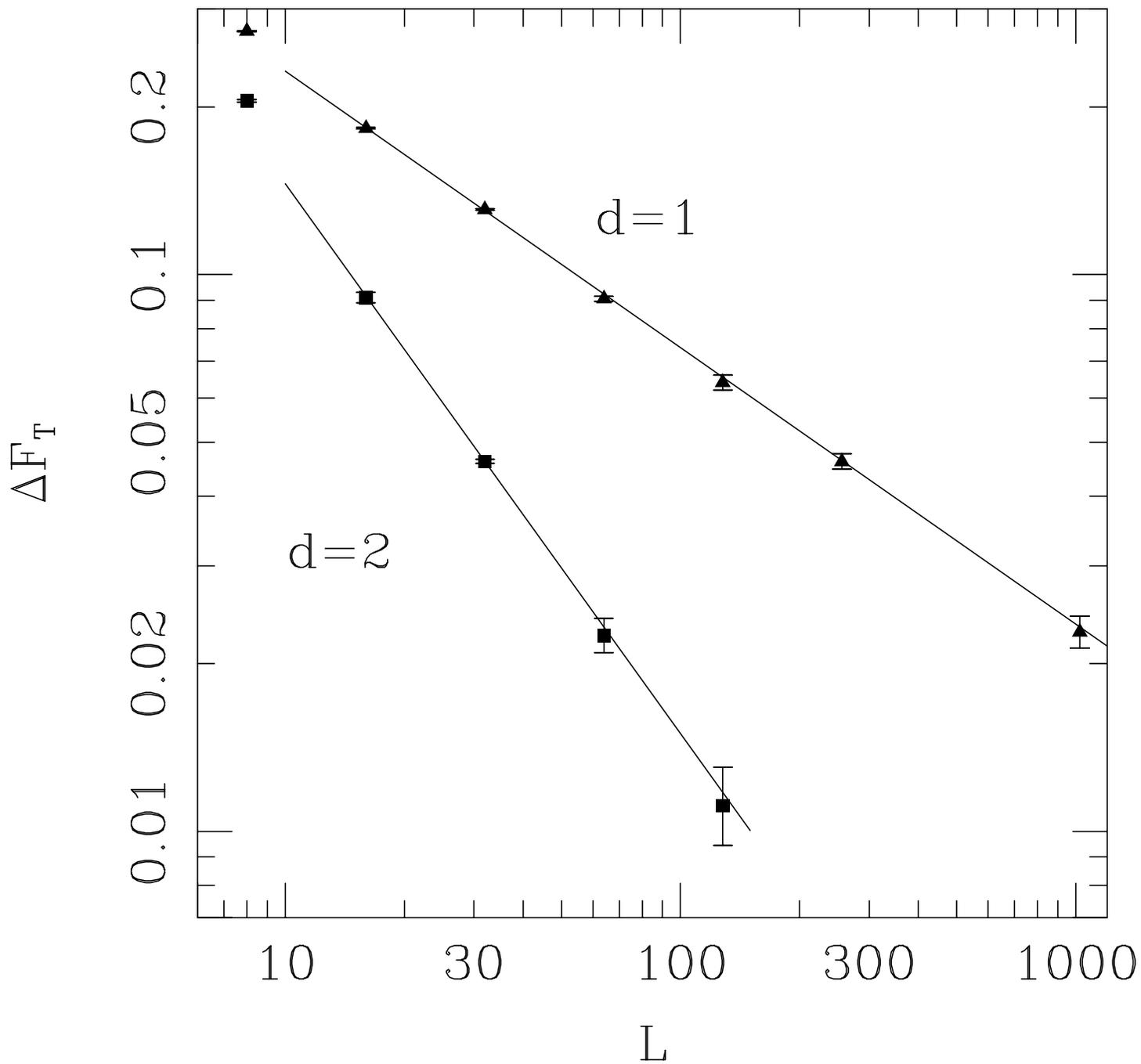
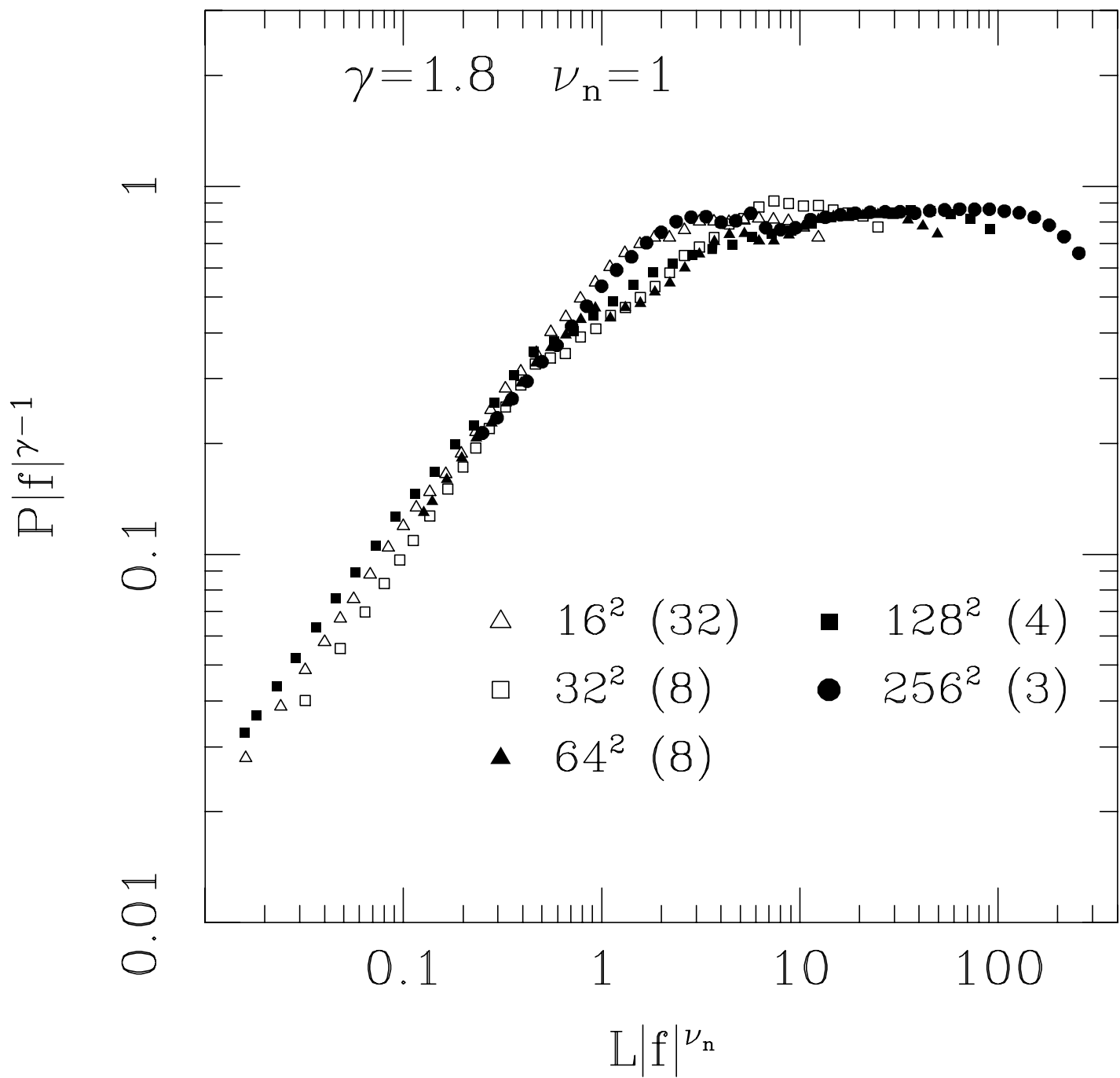
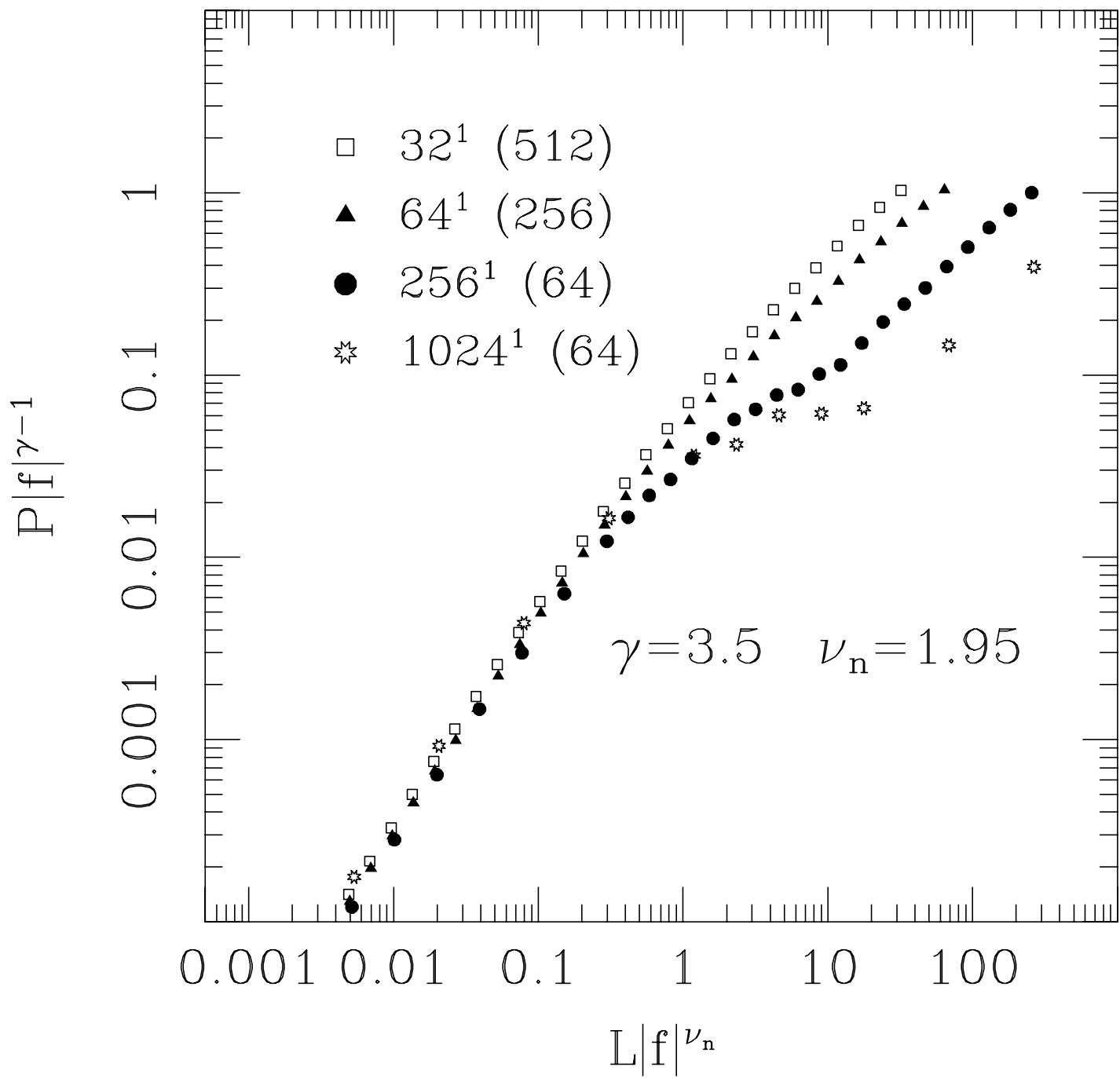
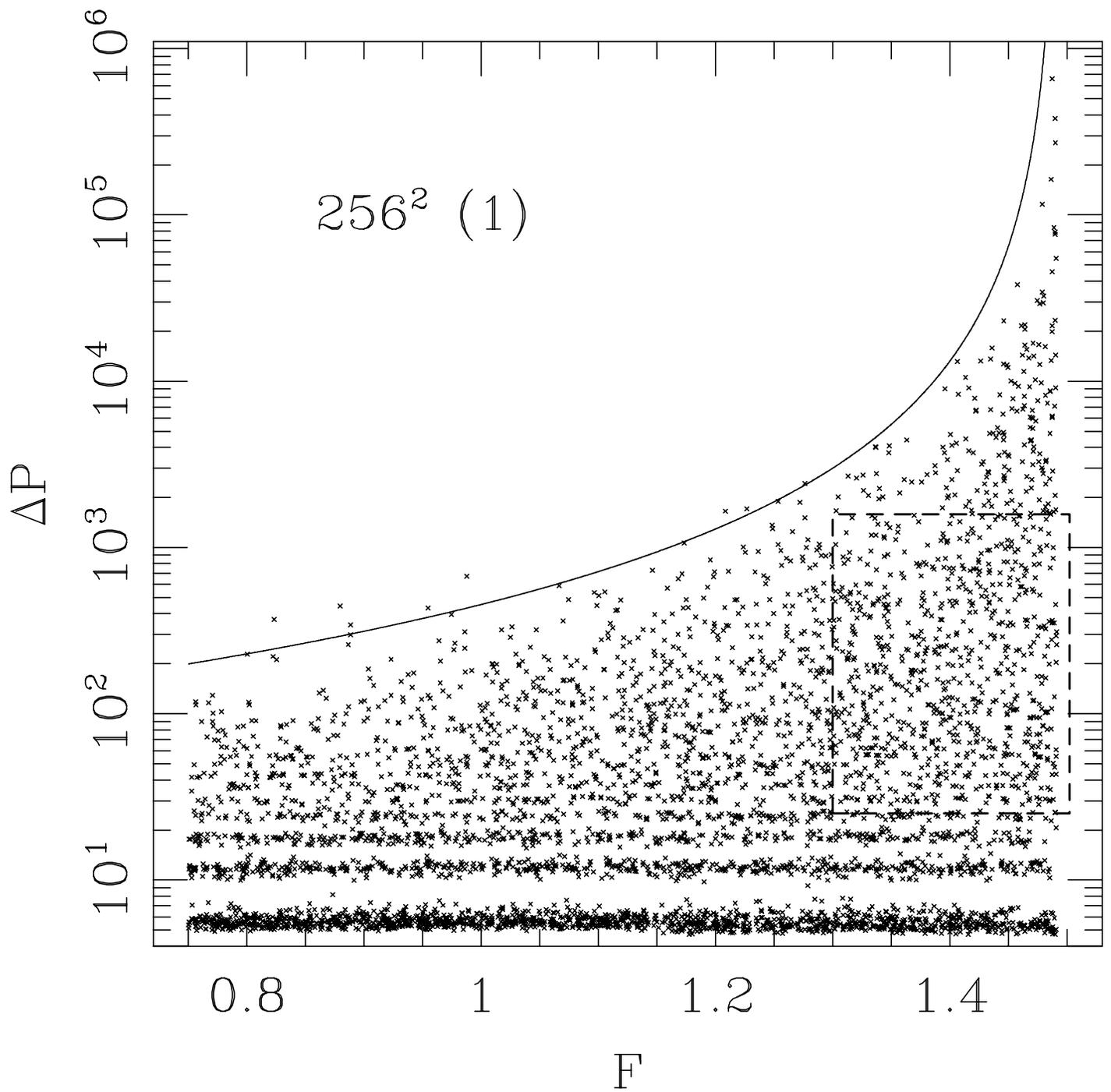
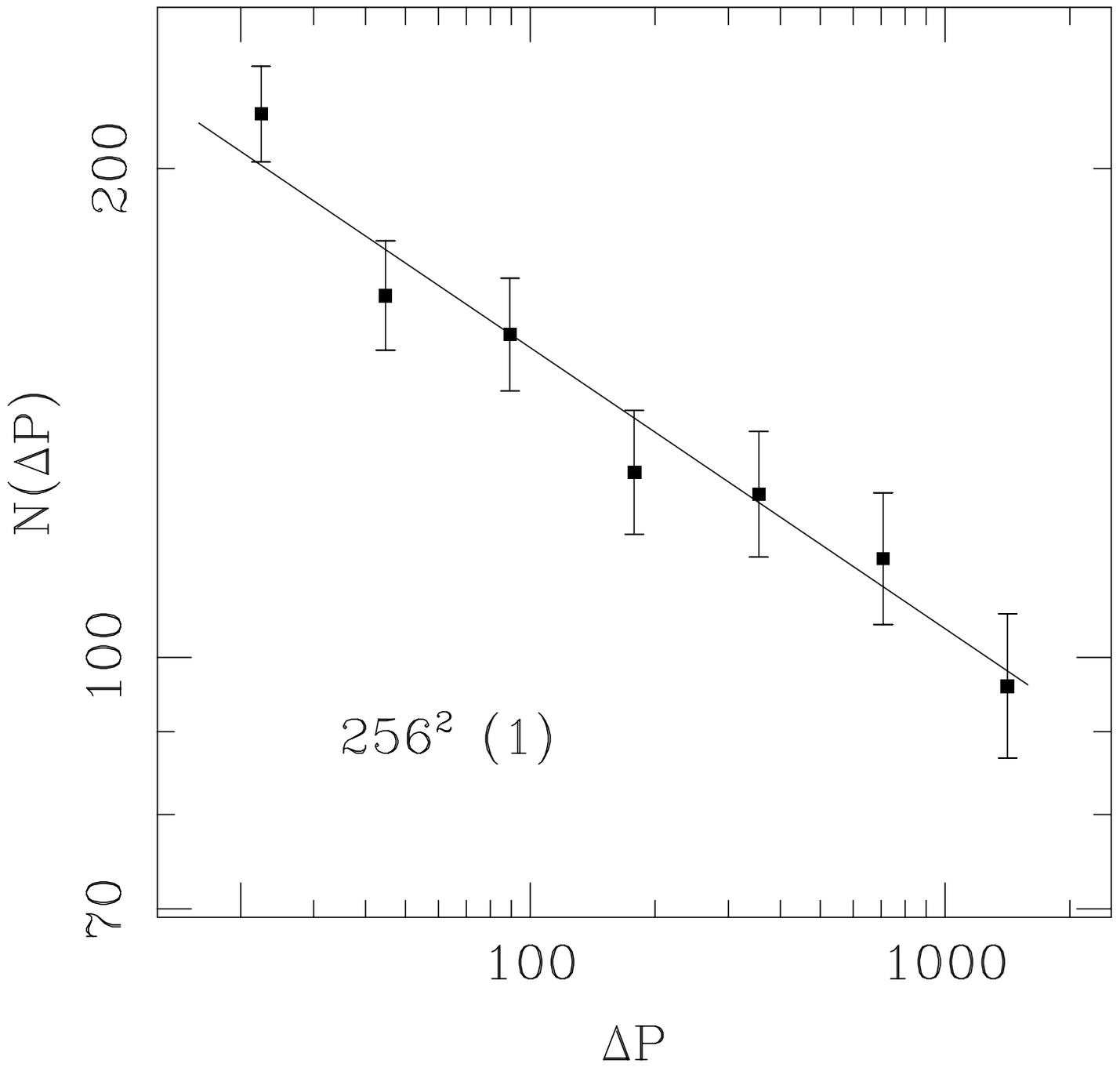
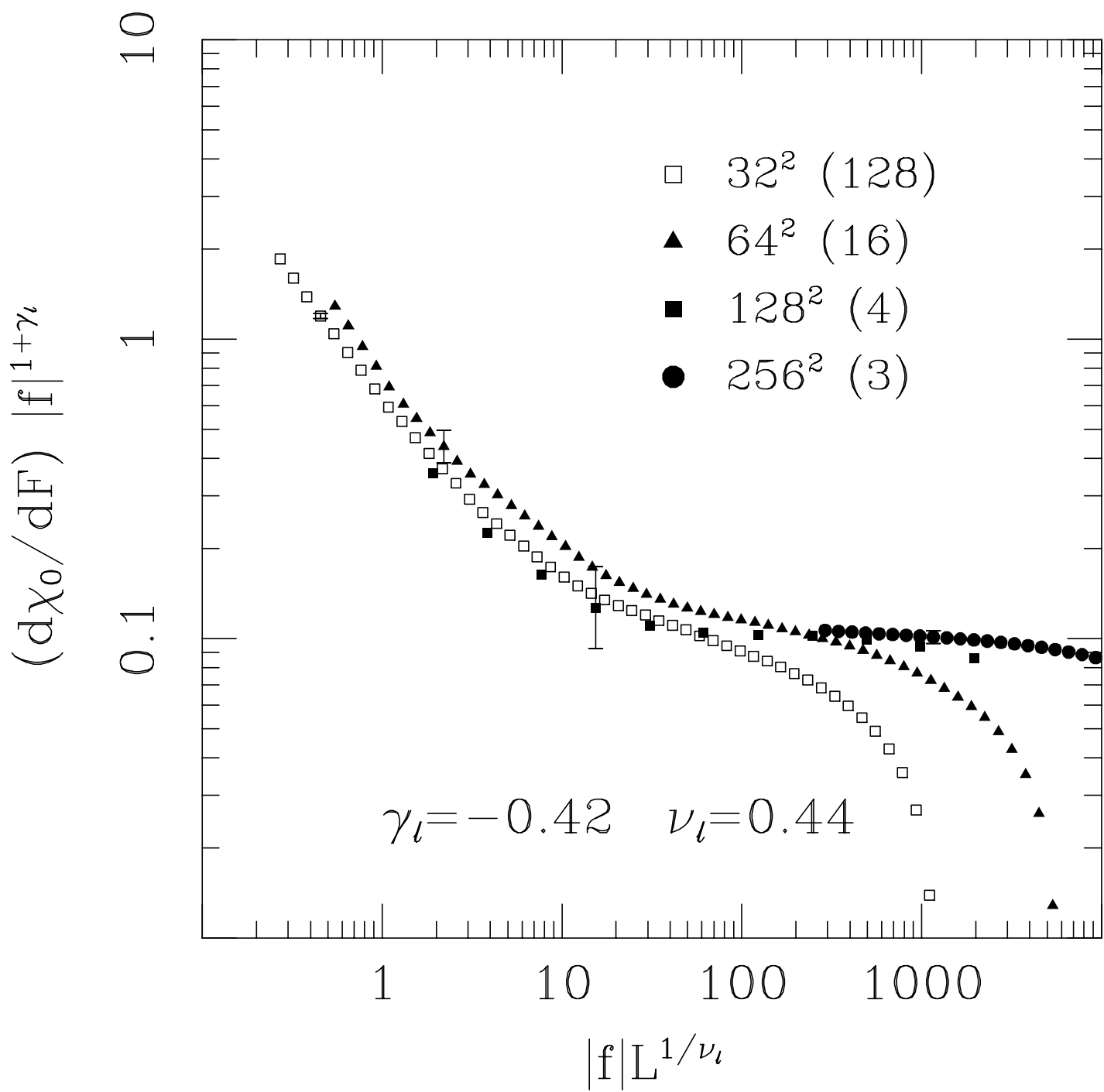
\end{document}